\documentclass[english,aps,a4paper,twocolumn,showpacs,amsfonts,amsmath,amssymb,amscd,nofootinbib]{revtex4-1}
\usepackage[english]{babel}
\usepackage{graphicx}
\usepackage[dvips]{color}
\usepackage{mathrsfs}
\usepackage{enumerate}
\usepackage{color}

\begin{document}

\def\Journal#1#2#3#4#5#6#7{{#1}, {\it #4} \textbf{#5}, #6 (#2).}
\def\Book#1#2#3#4#5{{#1}, {\it #3} (#4, #5, #2).}

\newcommand{\dd}{\mbox{d}}
\newcommand{\EE}{\mathbb{E}}
\newcommand{\NN}{\mathbb{N}}
\newcommand{\PP}{\mathbb{P}}
\newcommand{\RR}{\mathbb{R}}
\newcommand{\TT}{\mathbb{T}}
\newcommand{\ZZ}{\mathbb{Z}}
\newcommand{\uu}{\mathbf{1}}


\title{Monotonous continuous-time random walks with drift and stochastic reset events}
\author{Miquel Montero}
\email[E-mail: ]{miquel.montero@ub.edu}
\affiliation{Departament de F\'{\i}sica Fonamental, Universitat de Barcelona (UB), Mart\'{\i} i Franqu\`es 1, E-08028 Barcelona, Spain}
\author{Javier  Villarroel}
\email[E-mail: ]{javier@usal.es}
\affiliation{Facultad de Ciencias \& Instituto Universitario de F\'{\i}sica Fundamental y Matem\'aticas, Universidad de Salamanca, Plaza Merced s/n, E-37008 Salamanca, Spain}
\date{\today}
\begin{abstract}
In this paper we consider a stochastic process that may experience random reset events which bring suddenly the system to the starting value and analyze the relevant statistical magnitudes. We focus our attention on monotonous continuous-time random walks with a constant drift: the process increases between the reset events, either by the effect of the random jumps, or by the action of the deterministic drift. As a result of all these combined factors interesting properties emerge, like the existence|for any drift strength|of a stationary transition probability density function, or the faculty of the model to reproduce power-law-like behavior. General formulas for two extreme statistics, the survival probability and the mean exit time, are also derived. To corroborate in an independent way the results of the paper, Monte Carlo methods were used. These numerical estimations are in full agreement with the analytical predictions.
\end{abstract}
\maketitle
\section{Introduction}
Stochastic processes that incorporate some kind of mechanism which may instantaneously bring the {\it mathematical\/} system to an absorbing state are not new~\cite{Lefever,kt81}: The motivation may arise from many {\it real-life\/} systems which are threatened by the possibility of a sudden and severe decimation, usually with an external origin, e.g., the extinction of biological species due to some catastrophe, bankruptcy of industrial and financial firms or a major electric-power outage. 

Those situations where such a catastrophic event results in the disappearance of the system have attracted considerable interest and are well studied in the literature, both from a mathematical and physical perspective. Mathematical models that encompass this possibility are fairly developed: For diffusion processes (like the Brownian motion) the possibility  of extinction has been well described and several killing mechanisms have been considered in the literature. The introduction of an absorbing boundary at the origin, say, is the  simplest of them. In this situation the system vanishes once it hits the given level. However, the latter does not reflect all possible physical situations. Neither it does exhaust the possible behaviors at the boundary, which with more generality may display a combination of absorption, reflection, elastic, sticky and jump behavior|see \cite{Lefever,kt81} for the classification of boundaries for diffusion processes. A different mechanism, where boundary behavior plays no role, assumes that in any time interval $(t,t+\Delta t]$ the system has a probability of extinction $V(x)\Delta t$ where $x$ is the actual position and $V(x)$ the killing rate. Unlike what happened previously, in this setting extinction occurs in a sudden way, with no forewarning and may take place at any location $x$ and time $t$~\cite{kt81,Vtmp}.

In this paper we are interested in a related but nevertheless fairly different situation where the system is not extinguished but rather   is ``reborn from ashes." In such a case the (exogenous) event represents a reset in the system activity: the reboot of a computer, the 
neural spiking activity, or the ``back-to-square-one" order of certain board games.  The hallmark of  such a system is the possibility of a random restart, in opposition to random disappearance. Hence this novel behavior cannot be captured by considering alternative sorts of boundaries, like reflecting or sticky ones. Actually, here we are interested in  the opposite case where the triggering of resets is {\it totally independent of the underlying system\/}.

In spite of its obvious interest and significance, it appears that this situation has, until very recently, been virtually ignored in the literature. As far as we know, this possibility was first considered in the 1999 seminal work by Manrubia and Zanette~\cite{MZ99}. This paper analyzed a discrete-time stochastic multiplicative process with a {\it finite event space\/}, namely a Markov chain endowed with a restart mechanism. The second exception is due to  Evans and Majumdar~\cite{EM11a,EM11b} who in 2011, apparently unaware of the previous work of Manrubia and Zanette, considered the  diffusive evolution of a particle that may be randomly reset to its initial position.  
A remarkable conclusion, found both in~\cite{MZ99} and~\cite{EM11a}, was that the shut-off mechanism induces an algebraic decay for the tails of the corresponding probability density functions. {\it Power laws\/}, i.e., ``heavy-tailed" distributions, have been repeatedly observed in natural physical phenomena and advocated by several authors~\cite{Man,Ma,Stan,Sorn} but, despite this fact, power-law tails are by no means ubiquitous: most mathematical models ignore them in favor of Gaussian decay|see~\cite{BS} for a critical analysis of this dichotomy.

Following these ideas, in this paper we analyze the introduction of a restart mechanism in the framework of continuous-time random walks (CTRWs) with a constant drift and study the consequences that this carries. Thus, the dynamics of a such system involves three elements, namely the random jumps, the drift term and finally the resets.

If we deprive the system of the CTRW part but conserve the drift and the random reset mechanism we recover the classical {\it shot-noise process\/}, a well known model to describe sudden light emission with  current induction~\cite{Sch18,Rice}.

Conversely, pure CTRWs correspond to having only the first of these elements. The interest of these processes  in Physics  has been advocated specially by Montroll and Weiss~\cite{MW65,W94}. In Statistical Physics driftless CTRWs have been used to describe transport in disordered media~\cite{S74,ms84,Weiss-porra,Margolin1,Hu}, 
earthquake modeling~\cite{hs02,Me03}, random networks \cite{Be}, self-organized criticality in granular systems~\cite{Bo}, 
electron tunneling~\cite{Gu}, and distribution of matter in the universe~\cite{Os71}, just to name a few.
But CTRWs have been also used in the modeling of social systems, e.g., by giving a tick-by-tick description of financial markets~\cite{mmw03,mmp05,mmpw06,mpmlmm05}. A comprehensive review of CTRW applications in Finance and Economics can be found in~\cite{s06}. For a more general perspective on the issue see~\cite{Metzler,s04,m07}.

While not so popular in Physics, the CTRW with a drift incorporated plays a  fundamental role to model the cash flow at an insurance company~\cite{Gerbershiu,Dickson,lg04,zyl10}. Recently, we have shown that energy dissipation of optical beams in a self-defocusing medium like an optical fiber with random inhomogeneities is also described in terms of a CTRW with a drift~\cite{VM10,VM11}; see also~\cite{MV10}. Typically, in such a situation signal's losses may be significant and may require the incorporation of periodic all-optical amplifiers~\cite{ABCJS} to upgrade the signal strength to the initial amplitude. Thus, in this context, the exogenous reset mechanism proposed here might be incorporated to the model in a natural way.

The paper is organized as follows: In Sec.~\ref{Sec_process} we introduce the process under study, a CTRW  with drift on which a restart mechanism has been super-imposed. We suppose that resets are set off with no forewarning and independently  of the value of the underlying CTRW. The resulting process is  both rich and realistic enough to deserve important attention. In Sec.~\ref{Sec_transition}  we obtain an explicit analytic expression for the the transition probability or propagator of the system. It turns out that the full propagator can be related in a relatively simple way with the reset-free one.  We also find that as long as the reset mechanism is present, the asymptotic behavior (in either  time or space) shows some appealing features, like, in particular, the existence of a stationary distribution. We pinpoint conditions that guarantee the existence of {\it power-law tails\/} and characterize precisely the relevant exponent. 
Section~\ref{Sec_extreme} is devoted to the analysis of the properties of two extreme-value statistics, the survival probability and the mean exit time.  We give a closed formula for these statistics in terms of a Laplace transform. Conclusions are drawn in Sec.~\ref{Sec_conclusions}, where future perspectives are also sketched. We have left for the appendices some technical mathematical derivations.

\section{The process}
\label{Sec_process}
Our starting point is a random process $X(t)$ whose dynamics consists of a superposition of a constant rate motion along with sudden random increments $J_n$, the {\it jumps\/}, at random instants of time $t_n$. More concretely, if $T_0$ and $T_1$ are given times we suppose that for $T_0\leq t <T_1$, $X(t)$ can be expressed as
\begin{equation}
X(t)=\Gamma\times (t-T_0) +\sum_{n=-\infty}^{\infty} J_n \theta(t-t_n) \theta(t_n-T_0),
\label{process_T0}
\end{equation}
where $\Gamma\geq 0$ is the constant {\it velocity\/}, and $\theta(\cdot)$ is the Heaviside step function, that takes values $\theta(u)=1$ for $u\geq 0$ and zero otherwise. 

Furthermore, $t_n$ and $  J_n$ are the jump times and jump sizes while $\tau_n\equiv t_n-t_{n-1}$ are the time  {\it intervals\/} between two consecutive jumps, the ``waiting times."
We 
assume that $J_n$ and  $\tau_n $  define sequences of independent and identically distributed (i.i.d.) random variables,  described by the probability density functions (PDFs)  $h(\cdot)$ and $\psi(\cdot)$, respectively:
\begin{eqnarray*}
h(u)du &\equiv&
\PP\left\{u<J_n\leq u+du\right\},\\
\psi(\tau)d\tau &\equiv&
\PP\left\{\tau<\tau_n\leq \tau+d\tau\right\}.
\end{eqnarray*}
As we are interested in the analysis of {\it monotonous\/} processes, we will further assume that $h(u)=0$ for $u<0$ (whereas $\psi(\tau)=0$ for $\tau<0$ by definition.)

Up to the time $T_1$ Eq.~\eqref{process_T0} defines simply a CTRW with a constant drift $\Gamma$ like that considered in~\cite{MV10}. The difference with previous works arises at $t=T_1$, when the process  experiences a {\it reset event\/},~\footnote{Our mathematical model assumes that both jumps and reset events do 
suddenly modify the state of the process. While in some cases this 
instantaneous-response hypothesis will be an accurate description of  the system,  in others could only be considered as an acceptable proxy of reality.} i.e., we declare that $X(T_1)=0$ and that for $T_1\leq t <T_2$, $X(t)$ is given by
\begin{equation}
X(t)=\Gamma \times (t-T_1) +\sum_{n=-\infty}^{\infty} J_n \theta(t-t_n) \theta(t_n-T_1).
\label{process_T1}
\end{equation}
The same definition applies to $X(t)$ at $t=T_2, T_3, \ldots$, i.e., $X(T_m)=0$, where $T_m$ is an increasing sequence of random times. Thus in general for $T_{m}\leq t <T_{m+1}$, $m=0, 1, 2,\ldots$, we have
\begin{equation}
X(t)=\Gamma \times (t-T_m) +\sum_{n=-\infty}^{\infty} J_n \theta(t-t_n) \theta(t_n-T_m).
\label{process_Tm}
\end{equation}

\begin{figure}[htbp]
{
\includegraphics[width=0.9\columnwidth,keepaspectratio=true]{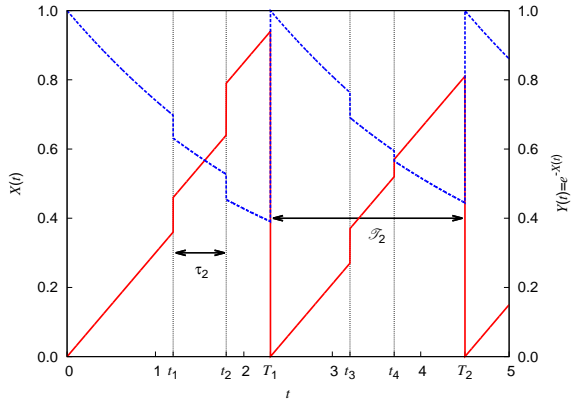}
}
\caption{(Color online) Sample path of the processes $X(t)$ and $Y(t)$. The solid (red) line represents a possible realization of the monotonous process $X(t)$ which grows linearly between the jump times $t_n$ and returns to the origin at the reset times $T_n$. The dashed (blue) line is a geometrical version of $X(t)$, $Y(t)=e^{-X(t)}$.} 
\label{Fig_sample}
\end{figure}

Note how past $T_1$ the initial evolution of $X(t)$, cf. Eq.~\eqref{process_T0}, is modified in an exogenous way. The novelty of this paper is the introduction of  this mechanism within the CTRW framework, which brings forward the possibility that the system be shut down and then restarted, see Fig.~\ref{Fig_sample}.

As regards the sequence of random resets $T_m$ we 
assume that the inter-event intervals $\mathscr{T}_m\equiv T_m-T_{m-1}$ are a set of i.i.d. random variables characterized by the PDF $\phi(\cdot)$,
\begin{equation*}
\phi(\tau)d\tau \equiv \PP\left\{\tau<\mathscr{T}_m\leq\tau+d\tau\right\}.
\end{equation*}
We also assume that the set $\left\{\mathscr{T}_m\right\}$ is independent of $\{\tau_n\}$ and $\{J_n\}$ and hence that {\it resets occur independently of the value of the underlying CTRW\/}. 

The process $X(t)$ thus modified is intrinsically different to the original one,~\footnote {In spite of this fact, one could argue that the long-time limit of the new system might follow by using an infinite combination of systems without resets. While in the analysis of some properties of the process this can be true,  see Eq.~\eqref{PDF_final} below,
it is unclear if this approach could be substantiated to produce concrete analytic results in the general case. We are grateful to an anonymous referee for this observation.} and, generically, is not a Markov process.
In fact, it is not either a compound renewal process: On the one hand, if $t_n$ is a given jump time, the evolution of $X(t)$ after that time depends on $X(t_n)$ but also on the previous reset time $T^*(t_n)\equiv T_{k}$, where $T_k < t_n\leq T_{k+1}$, for some $k$. On the other hand, if $T_m$ is a given reset time, the evolution of $X(t)$ after that time depends as well on the last jump time, which we denote by $t^*(T_m)\equiv t_l$, with $t_l<T_{m}\leq t_{l+1}$ for some value of $l$. In other words, in spite of the formal resemblance of Eqs.~\eqref{process_T0} and~\eqref{process_T1}, $X(T_0+\tau)$ and $X(T_1+\tau)$ have different laws and therefore
one cannot resort to the use of renewal equations in the analysis of this process for a general choice of $\psi(\cdot)$ and $\phi(\cdot)$, as is commonly done when studying CTRWs. 

In this paper we take the natural and usual choice that the waiting-time and reset densities  $\psi(\cdot)$ and $\phi(\cdot)$ are exponential functions
\begin{eqnarray}
\psi(\tau)&=&\lambda e^{-\lambda \tau},\label{Poisson_jumps}\\
\phi(\tau)&=&\Lambda e^{-\Lambda \tau},\label{Poisson_resets}
\end{eqnarray}
where $\lambda^{-1} $  and $\Lambda^{-1} $ are parameters of the distributions that respectively denote the mean value of $\tau_n$ and $\mathscr{T}_m$. 
Note that with this choice $X(t)$ is time-homogeneous and Markovian. Indeed,  if no resets are present $X(t)$ reduces to the well known compound Poisson process with drift which satisfies the above properties. (The interested reader can find the proof of this fact in  App.~\ref{App_Markov}, along with several global properties of this process.)

It is interesting to point out that in many cases the physically measurable process $Y(t)$ is not the given by Eq.~\eqref{process_Tm} but is rather constructed by multiplicative|instead of additive|increments. To accommodate this interpretation we simply consider that $X(t)$ is the (aggregated) rate at which the {\it physical observable\/}, $Y(t)$, varies. In such a case $X(t)$ and $Y(t)$ 
are related either by 
\begin{equation}
Y(t)=Y_0 \, e^{X(t)},\label{def_Yp}
\end{equation}
or by
\begin{equation}
Y(t)=Y_0 \, e^{-X(t)}.\label{def_Ym}
\end{equation}
This is the interpretation that we take in the rest of this paper where $X(t)$, defined by Eq.~\eqref{process_Tm}, is mostly considered for computational convenience, whereas the magnitude of interest is $Y(t)$. Let us enumerate a few examples:

In Finance, Eq.~\eqref{def_Yp} may represent the evolution of a savings account with a minimum balance level $Y_0$ (let us say, one cent) and a continuously compounded interest rate $\Gamma$, wherein sudden injections of capital are made by the owners at random times $t_n$, but which does not allow for partial withdrawals.

In Biology, Eq.~\eqref{def_Yp} may model the changes in the population of a species in some geographically-limited region, assuming  that it has an effective growth rate $\Gamma$ and is subjected to recurrent immigration and punctual massive emigrations: e.g., the case of many birds, swallows or flamingos among others. In this case the reset event does not imply the extinction of the species as a whole, but the return to the autochthonous level of population $Y_0$. 

In an optical context, the evolution of the energy $E(t)$ of a  signal in an optical fiber with random impurities is also described by a related model, as we  we have recently shown~\cite{VM10,VM11}. Here $E(t)\equiv E_0\, e^{-X(t)}$, where $X(t)$ is described by Eq.~\eqref{process_T0},  $\Gamma$ accounts for the constant damping rate of the fiber, while jump points $t_n$ signal the presence of random impurities in the fiber due to manufacture errors.  As a result of these combined effects, the initial signal will be degraded exponentially in time. The introduction of an ulterior reset mechanism, as we do here, could be used to describe the possibility that random (digital) amplifiers|at which the signal is amplified to the initial value, i.e., $E(T_n)=E_0$|are incorporated to the fiber to make up for the losses due to the damping and impurities.

Finally, we note that when $\lambda=0$, namely when only resets but no jumps are allowed, we 
have that Eq.~\eqref{def_Ym} leads to
\begin{equation*}
Y(t)=  Y_0 \times \sum_{m=-\infty}^\infty e^{-\Gamma (t-T_m)}
\theta(t-T_m)\theta(T_{m+1}-t).
\end{equation*}
Therefore, $Y(t)$ is the piece-wise exponential function $Y(t)= Y_0\, e^{-\Gamma (t-T_m)} \text{ for } T_{m}\leq t <T_{m+1}$. This is the  {\it shot-noise} process where shots are triggered with Poissonian rate and suffer an exponential decay after hitting the detector.  Shot noise has been used widely in Physics~\cite{Sch18,Rice} to describe sudden electron emission with induced currents, and Cherenkov radiation.

Thus, the  system considered here  is a natural generalization of classical physical models. In the rest of the paper we study the most relevant statistical  magnitudes.

\section{The transition PDF}
\label{Sec_transition}

We start with the analysis of the {\it propagator\/}, the transition probability density function of the process,
\begin{equation}
p_X(x,t;x_0,t_0)dx\equiv \PP\left\{\left. x< X(t) \leq x+dx\right| X(t_0)=x_0\right\},\\ \label{PDF_def} 
\end{equation}
where $t$ and $t_0$ are arbitrary~\footnote{We note that in the previous section we used $t_0$ to denote one of the jump times. This abuse of notation is not merely incidental. When the process is not time-homogeneous the renewal equations have a simpler structure if the time origin coincides with a jump.} with $t\geq t_0$. 

Note how in the definition of $p_X(x,t;x_0,t_0)$, the ``$X$" subscript indicates the random variable to which the transition PDF is associated; similarly, we denote by $p_Y(y,t;y_0,t_0)$ the corresponding probability for the physical process $Y(t)=e^{X(t)}$.~\footnote{We set $Y_0=1$ hereafter for computational convenience. In practical terms, this can always be achieved by a suitable choice of the physical units.}  However, since $p_Y$ follows trivially from $p_X$,
\begin{eqnarray}
p_Y(y,t;y_0,t_0)=\frac{1}{y}p_X(\log y,t;\log y_0;t_0), 
\label{p_Y}
\end{eqnarray}
we will mainly focus our attention on $p_X(x,t;x_0,t_0)$ in the sequel and, to simplify the notation, drop the ``$X$" subscript when there is no room for confusion.

Equation~(\ref{PDF_def}) intends to emphasize that, in principle, the propagator depends on both the starting point $x_0$ and the initial time $t_0$. However, the process must be time-homogeneous (since both the  jump and the reset mechanisms are Poissonian), and therefore
\begin{equation*}
p(x,t;x_0,t_0)=p(x,t-t_0;x_0,0)\equiv p(x,\tau;x_0).
\end{equation*}
By contrast, one does not expect the propagator to be translationally invariant because resets take the process to a fixed point, thereby breaking the spatial symmetry. This implies that the dependence on $x_0$ cannot be removed.

Let us derive first the integral equation that governs the evolution of $p(x,\tau;x_0)$. Recall that $t_0$ and $t$ (with $t_0\leq t $) are, respectively, the actual and a future given time instant, and that $\tau\equiv t-t_0$ is the associated time interval. In the following we will denote by $\tau'$ and $\tau''$ the time interval up to the first reset event and, respectively, the first jump|note that $\tau'$ and $\tau''$ are random variables while $\tau$ is just a number. The equation for the transition PDF can be built up by considering the three possible and mutually exclusive scenarios that appear depending on the relative values of the time interval up to the first reset event, $\tau'$, the time interval up to the first jump, $\tau''$, and $\tau$:
\begin{enumerate}[(i)]
\item 
There is neither a reset nor a jump in the time interval $\tau$, i.e., $\tau'>\tau$ and $\tau''>\tau$. In this case $X(\tau)=x_0+\Gamma \tau$.  
\item 
There is at least one reset in the time interval, and the first one takes place before any jump has occurred, $\tau'\leq\tau$ and $\tau''>\tau'$.  In this case the transition PDF after the reset must be $p(x,\tau-\tau';0)$ since system must reach $x$, starting from zero, in a lapse of time $\tau-\tau'$.
\item 
There is at least one jump in the time interval $(t_0,t]$, namely $\tau''\leq\tau$. This first jump (of size $u$) takes place before the first reset event, $\tau'>\tau''$. Right after the jump we have $X(\tau'')=x_0+\Gamma \tau'' +u$, and then the propagator is $p(x,\tau-\tau'';x_0+\Gamma \tau''+u)$.
\end{enumerate}
In view of all this $p(x,\tau;x_0)$ must satisfy the following renewal equation:
\begin{widetext}
\begin{eqnarray}
p(x,\tau;x_0)
&=&\int_\tau^{\infty}d\tau'\Lambda e^{-\Lambda \tau'}\int_\tau^{\infty}d\tau'' \lambda e^{-\lambda \tau''} \delta(x-x_0-\Gamma \tau) 
+\int_0^{\tau}d\tau'\Lambda e^{-\Lambda \tau'}\int_{\tau'}^{\infty}d\tau'' \lambda e^{-\lambda \tau''} p(x,\tau-\tau';0)\nonumber\\
&+&\int_0^{\tau}d\tau''\lambda e^{-\lambda \tau''}\int_{\tau''}^{\infty}d\tau' \Lambda e^{-\Lambda \tau'} \int_{0}^{\infty}du h(u)p(x,\tau-\tau'';x_0+\Gamma \tau''+u)\nonumber\\
&=& e^{-(\lambda+\Lambda) \tau} \delta(x-x_0-\Gamma \tau) +\int_0^{\tau}d\tau'\Lambda e^{-(\lambda+\Lambda) \tau'} p(x,\tau-\tau';0)
\nonumber\\
&+&\int_{0}^{\infty}du h(u) \int_0^{\tau}d\tau''\lambda e^{-(\lambda+\Lambda) \tau''} p(x,\tau-\tau'';x_0+\Gamma \tau''+u).
\label{TPDF}
\end{eqnarray}
\end{widetext}
The standard procedure for solving integral equations like~(\ref{TPDF}) is to resort to the use of some integral transformation, either the Laplace transform, the Fourier transform, or a combination of them. Here, since $x$ and $\tau$ are positive variables, the natural choice is to consider the Laplace transform in both arguments:
\begin{equation*}
\hat{\tilde{p}}(\omega,s;x_0)\equiv\int_0^{\infty}d\tau e^{-s \tau} \int_{0}^{\infty} dx p(x,\tau;x_0) e^{-\omega x},
\end{equation*}
where from now on the hat will denote the Laplace transform with respect to the {\it time\/} variable, and the tilde will denote the Laplace transform with respect to the {\it position\/} variable. Typically upon transformation the integral equation yields a purely algebraic equation whose solution can be (straightforwardly) found. In the present case, however, the double Laplace transform simplifies {\it but does not solve\/} the posed problem,
\begin{eqnarray}
\hat{\tilde{p}}(\omega,s;x_0)
&=&\frac{1}{\lambda +\Lambda+\Gamma \omega+s}e^{-\omega x_0} \nonumber\\
&+&\frac{\Lambda}{\lambda +\Lambda+s}\hat{\tilde{p}}(\omega,s;0)
\nonumber\\
&+&\int_0^{\infty}d\tau''\lambda e^{-(\lambda+\Lambda+s) \tau''} \nonumber\\ 
&\times &\int_{0}^{\infty}du h(u)\hat{\tilde{p}}(\omega,s;x_0+\Gamma \tau''+u),
\label{LTPDF}
\end{eqnarray}
since $\hat{\tilde{p}}(\omega,s;x_0)$ satisfies a new (although simpler) integral equation. 

Further  progress in the resolution of Eq.~\eqref{LTPDF} demands a careful analysis of the dependence of $p(x,\tau;x_0)$ on $x_0$ or, in  other words, determining which is the interplay between $x_0$, $x$ and $\tau$. In App.~\ref{App_ansatz} we prove that the solution to this integral equation is
\begin{eqnarray}
\hat{\tilde{p}}(\omega,s;x_0)&=&\left[\frac{\Lambda}{s}+e^{-\omega x_0}\right] \frac{1}{\lambda\left[1-\tilde{h}(\omega)\right] +\Lambda+\Gamma \omega+s},\nonumber\\ 
\label{p_sol}
\end{eqnarray}
where 
\begin{equation*}
\tilde{h}(\omega)\equiv\int_0^{\infty}du h(u) e^{-\omega u}.
\end{equation*}

The propagator $p(x,\tau;x_0)$ then follows by inversion of the Laplace transform of Eq.~\eqref{p_sol}. Letting $\Lambda \to 0$ one recovers $p_0(x,\tau;x_0)$, the propagator of a reset-free, monotonous CTRW with a drift, the process that follows if we let $T_1\to \infty$ in Eq.~\eqref{process_T0},
\begin{equation}
\hat{\tilde{p}}_0(\omega,s;x_0)=\lim_{\Lambda \to 0} \hat{\tilde{p}}(\omega,s;x_0)=\frac{e^{-\omega x_0}}{\lambda\left[1-\tilde{h}(\omega)\right] +\Gamma \omega + s}.
\label{p0_sol}
\end{equation}
It is noteworthy  that the opposite construction can also be carried out, namely $p(x,\tau;x_0)$ can be written in terms of $p_0(x,\tau;x_0)$. In physical space this relationship reads (see App.~\ref{App_ansatz})
\begin{eqnarray}
p(x,\tau;x_0)&=&\int_0^\tau d\tau^* \Lambda e^{-\Lambda \tau^*} p_0(x,\tau^*;0)\nonumber \\
&+&p_0(x,\tau;x_0)e^{-\Lambda \tau}.
\label{PDF_final}
\end{eqnarray}

Equation~\eqref{PDF_final}  is more informative from a structural point of view than Eq.~\eqref{p_sol} since it relates both propagators.  Actually the proof given in App.~\ref{App_Markov} on the Markovianity of $X(\tau)$ relies heavily on this fact. Nevertheless  we find it more convenient to use Eq.~\eqref{p_sol} as starting point for obtaining $p(x,\tau;x_0)$. In particular, one can go one step further and take the inverse Laplace transform in the $s$ variable of $\hat{\tilde{p}}(\omega,s;x_0)$ to obtain $\tilde{p}(\omega,\tau;x_0)$:
\begin{eqnarray}
\tilde{p}(\omega,\tau;x_0)&=&\frac{\Lambda\left[1-e^{-\left(\lambda\left[1-\tilde{h}(\omega)\right] +\Lambda+\Gamma \omega\right)\tau}\right]}{\lambda\left[1-\tilde{h}(\omega)\right] +\Lambda+\Gamma \omega} \nonumber\\
&+&e^{-\omega x_0 -\left(\lambda\left[1-\tilde{h}(\omega)\right] +\Lambda+\Gamma \omega\right)\tau}.
\label{p_sol_t}
\end{eqnarray}
An ulterior inverse Laplace transform in the $\omega$ variable could be given once the functional form of $\tilde{h}(\omega)$ is chosen.  Before examining further the general properties of the propagator we first consider a concrete example.

\subsection{Exponential jumps}
\label{Sub_exp}

Let us exemplify the general results in Eqs.~(\ref{p_sol})--(\ref{p_sol_t}) with the following choice for the jump density
\begin{equation}
h(u)=\gamma e^{-\gamma u},\, (u\geq0),
\label{h_exp}
\end{equation}
or equivalently
\begin{equation}
\tilde{h}(\omega)=\frac{\gamma}{\gamma +\omega}.
\label{tilde_h_exp}
\end{equation}
where $\gamma >0$ is a parameter. This election is both plausible and suitable, but even in this case the calculations are substantial. Since the purpose of this section is mainly illustrative we first consider the simpler case $\Gamma=0$.

Under these assumptions Eq.~\eqref{p_sol} reads
\begin{equation}
\hat{\tilde{p}}(\omega,s;x_0)=\left[\frac{\Lambda}{s}+e^{-\omega x_0}\right]\frac{\omega+\gamma}{\left(\lambda+\Lambda+s \right)\omega +\left(\Lambda+s\right)\gamma} ,\label{p_h_exp_ws}
\end{equation}
from which one can readily compute the inverse $\omega$--Laplace transform:
\begin{eqnarray}
\hat{p}(x,s;x_0)&=&\frac{1}{\lambda+\Lambda+s}\left[\frac{\Lambda}{s}\delta(x)+\delta(x-x_0)\right]\nonumber\\
&+&\frac{\Lambda }{s}\frac{\gamma \lambda}{\left(\lambda+\Lambda+s\right)^2}e^{-\frac{(\Lambda+s)\gamma x}{\lambda+\Lambda+s}}
\nonumber\\
&+&\frac{\gamma\lambda}{\left(\lambda+\Lambda+s\right)^2} e^{-\frac{\Lambda+s}{\lambda+\Lambda+s}\gamma (x-x_0)}\theta(x-x_0).\nonumber\\
\label{p_h_exp_s}
\end{eqnarray}
To completely recover the density one must, in addition, to perform the inverse Laplace transform in the $s$ variable, which is in turn more intricate.
After some calculations (see App.~\ref{app_p_x_tau}) we finally obtain 
\begin{eqnarray}
p(x,\tau;x_0)&=&\frac{\Lambda}{\lambda + \Lambda}\delta(x)\left[1-e^{-(\lambda+\Lambda)\tau}\right]\nonumber\\
&+&\delta(x-x_0)e^{-(\lambda+\Lambda)\tau}\nonumber\\
&+&\Lambda\int_0^\tau d\bar{\tau} e^{-(\lambda+\Lambda)\bar{\tau}-\gamma x}\sqrt{\frac{\gamma \lambda \bar{\tau}}{x}} I_1\left(2\sqrt{\lambda \bar{\tau}\gamma x}\right)\nonumber\\
&+&e^{-(\lambda+\Lambda)\tau-\gamma (x-x_0)}\sqrt{\frac{\gamma \lambda \tau}{x-x_0}} \nonumber\\
&\times& I_1\left(2\sqrt{\lambda \tau\gamma(x-x_0)}\right)\theta(x-x_0),
\label{p_x_tau}
\end{eqnarray}
where $I_1(\cdot)$ is the modified Bessel function of order 1.

Formula~(\ref{p_x_tau}) implies the existence of  a stationary PDF, which follows taking the limit $\tau\to\infty$ and computing the integral that appears in the third term.
Alternatively,  one finds in a straightforward way from~\eqref{p_h_exp_s} that
\begin{eqnarray}
p(x)&=&\lim_{\tau\rightarrow \infty}p(x,\tau;x_0)=\lim_{s\rightarrow 0}s \hat{p}(x,s;x_0)\nonumber\\
&=&\frac{\Lambda}{\lambda + \Lambda}\left[\delta(x)+\frac{\gamma\lambda}{\lambda+\Lambda}e^{-\frac{\Lambda\gamma x}{\lambda+\Lambda}}\right].
\label{p_h_exp}
\end{eqnarray}

Thus, the stationary distribution of the physically relevant quantity $Y(\tau)=e^{X(\tau)}$ 
follows from Eqs.~(\ref{p_Y}) and~(\ref{p_h_exp})
\begin{eqnarray}
p_Y(y)&=&\frac{1}{y}\frac{\Lambda}{\lambda + \Lambda}\left[\delta(\log y)+\frac{\gamma\lambda}{\lambda+\Lambda}e^{-\frac{\Lambda\gamma \log y}{\lambda+\Lambda}}\right]\nonumber\\
&=&\frac{\Lambda}{\lambda + \Lambda} \delta(y-1)+\frac{\gamma\lambda\Lambda}{\left(\lambda+\Lambda\right)^2}\frac{1}{y^{1+\alpha}}, 
\label{p_law}
\end{eqnarray}
with 
\begin{equation}
\alpha\equiv\frac{\gamma\Lambda}{\lambda+\Lambda}.
\label{alpha_exact}
\end{equation}


We close this section with some remarks concerning the stationary density for the general case $\Gamma>0$. The inclusion of a non-vanishing drift constant blurs somewhat the neat result in Eq.~(\ref{p_law}) with the emergence of two different exponents
\begin{equation}
p_Y(y)=\frac{\Lambda}{\Gamma(\alpha_+-\alpha_-)}\left[\frac{\alpha_+-\gamma}{y^{1+\alpha_+}}+\frac{\gamma-\alpha_-}{y^{1+\alpha_-}}\right],
\label{p_2law}
\end{equation}
with
\begin{equation}
\alpha_{\pm}\equiv\frac{\lambda +\Lambda+\Gamma \gamma}{2 \Gamma}\pm \frac{\sqrt{\left(\lambda +\Lambda+\Gamma \gamma\right)^2-4 \Gamma \gamma \Lambda }}{2 \Gamma}.
\label{alpha_pm}
\end{equation}
Since one has $\alpha_+>\gamma>\alpha_->0$, the bigger parameter $\alpha_+$ dominates the behavior of $p_Y(y)$ for $y\sim 1$ whereas the smaller one, $\alpha_-$, determines its asymptotic decay and defines the power-law exponent of the process. The critical value of $y$ that separates these two regimes is
\begin{equation*}
y_{\rm c}=\left(\frac{\alpha_+-\gamma}{\gamma-\alpha_-}\right)^{\frac{1}{\alpha_+-\alpha_-}}.
\end{equation*}
Therefore, when the jumps are exponentially distributed, we find a power-law decay for $p_Y(y)$ for any relative value of the parameters.

\begin{figure}[htbp]
{
\includegraphics[width=0.9\columnwidth,keepaspectratio=true]{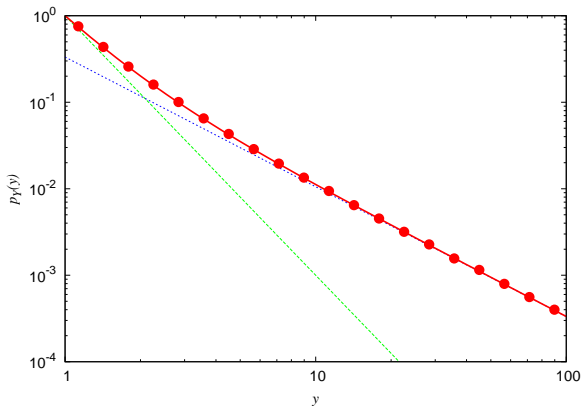}
}
\caption{(Color online) Power-law decay in $p_Y(y)$. 
The solid (red) line corresponds to the stationary propagator, Eq.~\eqref{p_2law}, whereas the dashed (green) line and the dotted (blue) line correspond to a decay rate driven exclusively by $\alpha_+$ or $\alpha_-$ respectively. Superimposed to the exact result, and in a good agreement with it, we can find (circles) a numerical estimation of the stationary PDF obtained through Monte Carlo simulation.}
\label{Fig_p2law}
\end{figure}

To test the goodness of the previous analytical derivations we have resorted to numerical methods. We have simulated the evolution of  $1,000,000$ different realizations of the process $Y(\tau)$: the relevant  PDF is then derived based on the relative frequency of the outcomes.  We plot this PDF for $\tau=100$, a quantity that is much larger than any  characteristic timescale of the process, since we have set $\Gamma=2.0$, $\lambda=1.0$, $\Lambda=2.0$, and $\gamma=1.0$, in arbitrary units.  For such large values of time, the process is to all effects within the stationary regime, as Fig.~\ref{Fig_p2law} confirms.

Figure~\ref{Fig_p2law} also shows the coexistence of the two different exponents in the decay rate of the stationary PDF, $\alpha_+$ and $\alpha_-$. With the choice of the parameters reported above $\alpha_+=2.0$, $\alpha_-=0.5$, and $y_{\rm c}=\sqrt{8}$.

\subsection{Some remarkable general properties}

We have seen in the previous section that if the jump density is exponentially distributed  the system tends for long times to a stationary state where the stationary density has a power law decay. We now  show that {\it this behavior is not  restricted to this particular case but occurs under fairly general conditions and election of jump density $h(\cdot)$}.

To this end note that in the Laplace domain $\tilde{p}(\omega,\tau;x_0)$ 
is a well behaved function of $\omega$ provided $\tilde{h}(\omega)$ presents no pathologies. Nevertheless, $\tilde{h}(\omega)\to 0$ as $\omega\to\infty$ and hence the inversion to physical space must be done in a generalized distributional sense that requires some care: indeed the last term of Eq.~(\ref{p_sol_t}) has no inverse Laplace transform for $x\geq x_1\equiv x_0+\Gamma \tau$ and a delta contribution appears at that point. This anomaly reflects that there is a finite probability that the process arrives to $x_1$ before any reset or jump has occurred:    
\begin{equation}
\PP\left\{\left. X(\tau)=x_1\right| X(0)=x_0\right\}= e^{-(\Lambda+\lambda)\tau}.
\label{delta_Gamma}
\end{equation}
When $\Gamma=0$ there is, in addition, a finite chance of finding the process at $x=0$ after time $\tau$, irrespective of the initial point. This probability has its origin in the first term of Eq.~(\ref{p_sol_t}), and amounts
\begin{equation}
\PP\left\{\left. X(\tau)=0\right| X(0)=x_0\right\}=\frac{\Lambda}{\Lambda+\lambda}\left[1- e^{-(\Lambda+\lambda)\tau}\right].
\label{delta_no_Gamma}
\end{equation}
Remember how these two delta terms have previously appeared in Eq.~\eqref{p_x_tau}. Note also that when $x_0=\Gamma=0$ the probabilities in Eqs.~(\ref{delta_Gamma}) and~(\ref{delta_no_Gamma}) will simply add together thus yielding
\begin{equation*}
\PP\left\{\left. X(\tau)=0\right| X(0)=0\right\}=\frac{\Lambda+\lambda e^{-(\Lambda+\lambda)\tau}}{\Lambda+\lambda}.
\end{equation*}

It is also remarkable that the introduction of resets makes the system recurrent and ergodic 
with a stationary density that follows from equations~(\ref{p_sol}) and~(\ref{p_sol_t}):
\begin{eqnarray}
\tilde{p}(\omega)&\equiv&\lim_{\tau\rightarrow \infty}\tilde{p}(\omega,\tau;x_0)=\lim_{s\rightarrow 0} s \hat{\tilde{p}}(\omega,s;x_0)\nonumber \\
&=&
\frac{\Lambda}{\lambda \left[1-\tilde h(\omega)\right]+\Lambda+\Gamma \omega}.
\label{stat_PDF}
\end{eqnarray}
This ergodicity could be expected on physical grounds since the incorporation of this reset mechanism guarantees that the system will not be driven too far off from the origin, no matter how strong the drift and jump mechanisms are. We elaborate on these properties in App. ~\ref{App_Markov}. Note that while the limit does not depend on the initial condition $x_0$, as expected, {\it it still depends on the drift velocity\/} $\Gamma$, a fact that is not obvious from Eq.~(\ref{TPDF}), where $x_0$ and $\Gamma$ always appear side by side.

In particular, we can use expression~\eqref{stat_PDF} to compute the moments of the stationary distribution for a general choice of $h(\cdot)$. Thus, if we call $\mu_n$ the $n$th-order moment of the jump-size PDF $h(\cdot)$ we will have, for instance,
\begin{eqnarray*}
\lim_{\tau\to\infty}\EE\left[X(\tau)\right]&=&\frac{\Gamma + \lambda\mu_1}{\Lambda} ,\\
\lim_{\tau\to\infty}\EE\left[X(\tau)^2\right]&=&\frac{\lambda}{\Lambda} \mu_2+\frac{2\left(\Gamma + \lambda\mu_1\right)^2}{\Lambda^2},
\end{eqnarray*}
where $\EE[\cdot]$ denotes the expectation of its argument.


We are particularly interested in studying the asymptotic behavior of $p_X(x)$ for large value of their arguments. By virtue of the Tauberian theorem this amounts to studying the limit of Eq.~(\ref{stat_PDF}) when $\omega\rightarrow
0$ and $h(\omega)\rightarrow 1$. Depending on the magnitude of $\Lambda$ different scenarios appear. Under the assumption that the reset rate is much larger than the jump rate, $\Lambda\gg \lambda$, and the {\it velocity\/}, $\Lambda\gg \Gamma$, we expand Eq.~(\ref{stat_PDF}) as
\begin{eqnarray*}
\tilde{p}(\omega)&=&\frac{1}{1+\frac{\lambda}{\Lambda} \left[1-\tilde h(\omega)\right]+\frac{\Gamma }{\Lambda}\omega}\nonumber \\
&\sim&1-\frac{\lambda}{\Lambda} -\frac{\Gamma}{\Lambda}\omega +\frac{\lambda}{\Lambda}\tilde h(\omega).
\end{eqnarray*}
This expression leads to
\begin{eqnarray*}
p(x)\sim\frac{\lambda}{\Lambda} h(x),
\end{eqnarray*}
where we have rejected the terms that do not contribute to the asymptotic behavior. 

More compelling is the behavior in the opposite case when the above relations are inverted. Concretely, if  the jump density has finite mean $\mu_1$ then for small values of $\omega$ we can
approximate 
\begin{equation*}
\tilde h(\omega)\sim 1-\mu_1 \omega +o(\omega),
\end{equation*}
 and hence  Eq.~(\ref{stat_PDF}) reads:
\begin{eqnarray}
\tilde{p}(\omega)\sim \frac{\Lambda}{\left[\lambda \mu_1 +
\Gamma\right]\omega+\Lambda}. \label{p_law_gen_a}
\end{eqnarray}
Assume in addition that  the ratio
\begin{equation}
\frac{\Lambda}{\lambda \mu_1 + \Gamma}\sim O(\omega),
\label{beta_cond}
\end{equation}
i.e., it is at least as small as $\omega$ is.  In this case the Laplace inversion yields for large $x$
\begin{eqnarray}
p(x)\sim \frac{\Lambda}{\lambda \mu_1 + \Gamma}
e^{-\frac{\Lambda}{\lambda \mu_1 + \Gamma}x}. \label{p_law_gen_b}
\end{eqnarray}

It is interesting to observe the emergence of different regimes in $p_Y(y)$ as well, as we increase the reset parameter $\Lambda$.  Setting $\Lambda=0$ (i.e., if we freeze the reset mechanism) the stationary density disappears as befits an increasing unbounded process. When we introduce a small reset parameter $\Lambda $, the stationary density $p_Y(y)$ exists and if condition~\eqref{beta_cond} is satisfied it  will show a heavy-tail decay
\begin{eqnarray}
p_Y(y)\sim \frac{\beta}{y^{1+\beta}}, \label{p_law_gen_c}
\end{eqnarray}
with power-law exponent $\beta$,
\begin{equation}
\beta\equiv\frac{\Lambda}{\lambda \mu_1 + \Gamma}.
\label{alpha_gen}
\end{equation}
Note that this proves that under the above assumptions the stochastic process $Y(\tau)$ is  {\it ergodic with a stationary density that shows a power-law decay\/}. The critical exponent
depends on a sophisticated combination of all the parameters that characterizes the model: the jump and reset rates, the drift {\it velocity\/} and the typical jump sizes. The result is similar to that
reported in~\cite{MZ99}.

As one keeps increasing the reset rate so that $\Lambda \gg \lambda$ and $\Lambda\gg \Gamma$, the stationary density is dominated by the functional form of the jump density, which may or may not decay rationally:
\begin{eqnarray*}
p_Y(y)\sim\frac{1}{y} h\left(\log y\right).
\end{eqnarray*}

Finally,  for ever larger values of  $\Lambda $ the pure density part  is dimmed while the delta term takes over and dominates the stationary probability distribution, a result which is easy to understand from a physical point of view.

We finish this discussion by noting  that, in the given ranges, these expressions reduce to the results obtained in the previous section: one just has to recall that 
from Eq.~\eqref{h_exp} one has $\mu_1=\gamma^{-1}$.

\section{Extreme-event statistics}
\label{Sec_extreme}

In this Section we analyze some statistical properties of extreme events of the process $X(\tau)$.

We are concerned with the survival probability (SP) of the process, $\mathcal{P}_{[a,b]}(\tau;x_0)$,  
\begin{equation}
\mathcal{P}_{[a,b]}(\tau;x_0)\equiv \PP\left\{\left. a\leq X(\bar{\tau}) \leq b, \bar{\tau} \leq \tau \right| X(0)=x_0\right\},
\end{equation}
namely, the probability that the process which is initially in $x_0$ does not leave the region $[a,b]$, $a\leq x_0\leq b$, before time $\tau$. In other words, if we denote by $\mathcal{T}$ the first time the process exits $[a,b]$ starting from $x_0$, 
\begin{equation}
\mathcal{T}\equiv\min \left\{\tau: X\left(\tau\right)\notin[a,b]| X(0)=x_0 \right\},
\label{Exit_time_def}
\end{equation}
then the SP 
is simply the probability that $\mathcal{T}>\tau$, that is
\begin{equation}
\PP\left\{\mathcal{T}\leq\tau\right\}=1-\mathcal{P}_{[a,b]}(\tau;x_0).
\label{survival}
\end{equation}

A related magnitude of interest is the mean exit time (MET), the expected value of $\mathcal{T}$:
\begin{equation}
\TT_{[a,b]}(x_0)\equiv \EE\left[\mathcal{T}\right].
\end{equation}
Even though these two magnitudes are meaningful for any choice of the interval $[a,b]$ we are primarily interested in the case $a=0$, $b>0$: By construction the process will tend to reach the upper boundary $b$ steadily, but occasionally it returns to the lower boundary $a=0$. 

To ease the notation in the sequel we drop the subscript in $x_0$, and assume that the process is initially at $x$. In order to get an equation for $\mathcal{P}_{[0,b]}(\tau;x)$ we will resort to renewal arguments, very similar to those enumerated in Sec.~\ref{Sec_transition}, but with the additional requirement that the process does not  cross the upper boundary at any time. With the same notation as in Sec.~\ref{Sec_transition} we will explore the three non-overlapping situations :
\begin{enumerate}[(i)]
\item There is neither a jump nor a reset in the time interval $\tau$, $\tau'>\tau$ and $\tau''>\tau$, and the elapsed time is not long enough to reach the boundary by the effect of the drift, $\tau\leq (b-x)/\Gamma$. In this case the process will survive with certainty.  
\item There is at least one reset event at $\tau'$ before the first jump, $\tau' \leq \tau$, $\tau'<\tau''$. If $\tau'> (b-x)/\Gamma$ the process does not survive, otherwise the survival probability is $\mathcal{P}_{[0,b]}(\tau-\tau';0)$.
\item There is at least one jump before the first reset at $\tau''$, $\tau''<\tau'$, $\tau'' \leq \tau$. After this event there is survival probability $\mathcal{P}_{[0,b]}(\tau-\tau'';x+\Gamma \tau'+u)$ only if $\tau''\leq (b-x)/\Gamma$ and the jump size $u$ is smaller than $b-x-\Gamma \tau''$.
\end{enumerate}
These three possible contingencies lead to the following integral equation:
\begin{widetext}
\begin{eqnarray}
\mathcal{P}_{[0,b]}(\tau;x)&=&\int_\tau^{\infty}d\tau'\Lambda e^{-\Lambda \tau'} \int_\tau^{\infty}d\tau'' \lambda e^{-\lambda \tau''}\theta(b-x-\Gamma \tau)
\nonumber\\
&+&\int_0^{\tau}d\tau'\Lambda e^{-\Lambda \tau'}\int_{\tau'}^{\infty}d\tau'' \lambda e^{-\lambda \tau''} \mathcal{P}_{[0,b]}(\tau-\tau';0)\theta(b-x-\Gamma \tau')
\nonumber\\
&+&\int_0^{\tau}d\tau''\lambda e^{-\lambda \tau''}\int_{\tau''}^{\infty}d\tau' \Lambda e^{-\Lambda \tau'} \int_{0}^{b-x-\Gamma \tau''}du h(u)\mathcal{P}_{[0,b]}(\tau-\tau'';x+\Gamma \tau''+u) \theta(b-x-\Gamma \tau'')
,\nonumber\\
&=& e^{-(\lambda+\Lambda) \tau} \theta(b-x-\Gamma \tau)
+\int_0^{\tau}d\tau'\Lambda e^{-(\lambda+\Lambda) \tau'}\mathcal{P}_{[0,b]}(\tau-\tau';0)\theta(b-x-\Gamma \tau')
\nonumber\\
&+&\int_0^{\tau}d\tau''\lambda e^{-(\lambda+\Lambda) \tau''} \int_{0}^{b-x-\Gamma \tau''}du h(u)\mathcal{P}_{[0,b]}(\tau-\tau'';x+\Gamma \tau''+u)\theta(b-x-\Gamma \tau''). 
\end{eqnarray}
\end{widetext}
Next we will consider the Laplace transform of the SP with respect to the time variable, 
\begin{equation}
\hat{\mathcal{P}}_{[0,b]}(s;x)\equiv\int_0^{\infty}d\tau \mathcal{P}_{[0,b]}(\tau;x) e^{-s \tau},
\label{LSP_def}
\end{equation}
and obtain
\begin{eqnarray}
\hat{\mathcal{P}}_{[0,b]}(s;x)&=&\frac{1+\Lambda \hat{\mathcal{P}}_{[0,b]}(s;0) }{\lambda+\Lambda+s}\left[1- e^{-\frac{\lambda+\Lambda+s}{\Gamma}(b-x)}\right] 
\nonumber\\
&+&\frac{\lambda}{\Gamma}\int_0^{b-x}dz e^{-\frac{\lambda+\Lambda+s}{\Gamma}(b-x-z)} \nonumber\\
&\times&\int_{0}^{z}du h(u)\hat{\mathcal{P}}_{[0,b]}(s;b-z+u).
\label{Full_PL}
\end{eqnarray}
Note that in this case we cannot perform the Laplace transform with respect to $x$ in a straightforward way because 
$x$ is restricted to the interval $[0,b]$. In the following section we show how to overcome this limitation.

We also remind (see App.~\ref{App_s_is_0}) that setting $s=0$ in Eq.~(\ref{Full_PL}) the corresponding integral equation for the mean exit time of the process out of the interval, $\TT_{[0,b]}(x)$, is directly obtained:~\footnote{One can alternatively obtain
this equation by means of the same kind of arguments that led to Eq.~(\ref{Full_PL}), see, e.g., \cite{mmp05,MV10}.}
\begin{eqnarray}
\TT_{[0,b]}(x)&=&\frac{1+\Lambda \TT_{[0,b]}(0) }{\lambda+\Lambda}\left[1- e^{-\frac{\lambda+\Lambda}{\Gamma}(b-x)}\right] 
\nonumber\\
&+&\frac{\lambda}{\Gamma}\int_0^{b-x}dz e^{-\frac{\lambda+\Lambda}{\Gamma}(b-x-z)} \nonumber \\
&\times&\int_{0}^{z}du h(u)\TT_{[0,b]}(b-z+u).
\label{Full_TL}
\end{eqnarray}

\subsection{The general solution}
Here we consider the solution to Eq.~(\ref{Full_PL}) with arbitrary choice of the jump size PDF $h(\cdot)$. To this end we have to find the general solution of the allied integral equation 
\begin{eqnarray}
\hat{\mathcal{F}}(s;z)&=&\frac{1+\Lambda \hat{\mathcal{A}}(s)}{\lambda+\Lambda+s}\left[1- e^{-\frac{\lambda+\Lambda+s}{\Gamma}z}\right] 
\nonumber\\
&+&\frac{\lambda}{\Gamma}\int_0^z dz' e^{-\frac{\lambda+\Lambda+s}{\Gamma}(z-z')}\nonumber\\
&\times&\int_{0}^{z'}du h(u)\hat{\mathcal{F}}(s;z'-u),
\label{Full_FL}
\end{eqnarray}
for $z\geq 0$, with $\hat{\mathcal{A}}(s)$ an arbitrary function of $s$.~\footnote{Note that in Eq.~\eqref{Full_PL} $s$ is a parameter and hence to all effects $\hat{\mathcal{A}}(s)$ plays the role of a constant.} Notice that one recovers $\hat{\mathcal{P}}_{[0,b]}(s;x)$ via $\hat{\mathcal{P}}_{[0,b]}(s;x)=\hat{\mathcal{F}}(s;b-x)$ and requiring that $\hat{\mathcal{A}}(s)=\hat{\mathcal{P}}_{[0,b]}(s;0)$. The solution of $\hat{\mathcal{F}}(s;z)$ for values larger than $b$, i.e., when $x<0$, is mathematically meaningful but has no physical significance.

Expression~(\ref{Full_FL}) is now well suited to be Laplace transformed in the $z$ variable as well,
\begin{equation*}
\hat{\tilde{\mathcal{F}}}(s;\omega)\equiv \int_0^{\infty}dz \hat{\mathcal{F}}(s;z) e^{-\omega z},
\end{equation*}
yielding in this way
\begin{eqnarray}
\hat{\tilde{\mathcal{F}}}(s;\omega)&=&\frac{1}{\omega}\frac{1+\Lambda \hat{\mathcal{A}}(s)}{\lambda+\Lambda+\Gamma \omega+s}\nonumber\\
&+&\frac{\lambda}{\lambda+\Lambda+\Gamma \omega+s} \tilde{h}(\omega) \hat{\tilde{\mathcal{F}}}(s;\omega).
\end{eqnarray}
Then we can obtain a closed solution of the problem for any functional form of $h(\cdot)$ in the Laplace-Laplace domain:
\begin{eqnarray}
\hat{\tilde{\mathcal{F}}}(s;\omega)=\frac{1}{\omega}\frac{1+\Lambda \hat{\mathcal{A}}(s)}{\lambda \left[1-\tilde{h}(\omega)\right]+\Lambda+\Gamma \omega+s}.
\label{Closed_P}
\end{eqnarray}

\subsection{Exponential jumps}
Let us exemplify the general result in Eq.~(\ref{Closed_P}) with the exponential case, introduced earlier in Eq.~(\ref{h_exp}). 

Let us begin by writing down the explicit expression for $\hat{\tilde{\mathcal{F}}}(s;\omega)$,
\begin{eqnarray}
\hat{\tilde{\mathcal{F}}}(s;\omega)&=&\frac{\gamma +\omega}{\omega}\frac{1+\Lambda \hat{\mathcal{A}}(s)}{\Gamma \omega^2 +\left(\lambda +\Lambda+s+\Gamma \gamma\right) \omega + \gamma\left(\Lambda + s\right)}\nonumber\\
&=&\frac{1+\Lambda \hat{\mathcal{A}}(s)}{\Delta(s)} \frac{1}{\omega} \left[\frac{\alpha_+(s)-\gamma}{\omega+\alpha_+(s)}+\frac{\gamma-\alpha_-(s)}{\omega+\alpha_-(s)}\right],\nonumber \\
\label{Exp_a}
\end{eqnarray}
where
\begin{eqnarray}
\alpha_{\pm}(s)&\equiv&\frac{\lambda +\Lambda+s+\Gamma \gamma}{2 \Gamma}\pm \frac{\Delta(s)}{2 \Gamma},\\
\Delta(s)&\equiv&\sqrt{\left(\lambda +\Lambda+s+\Gamma \gamma\right)^2-4 \Gamma \gamma \left(\Lambda + s\right)}.
\end{eqnarray}
Note that the $\alpha_{\pm}$ defined above in Eq.~(\ref{alpha_pm}) corresponds simply to $\alpha_{\pm}=\alpha_{\pm}(0)$.

The $\omega$--Laplace inversion of Eq.~(\ref{Exp_a}) is lengthy but straightforward,
\begin{eqnarray}
\hat{\mathcal{F}}(s;z)
&=&\frac{1+\Lambda \hat{\mathcal{A}}(s)}{\Delta(s)}\left[\frac{\alpha_+(s)-\gamma}{\alpha_+(s)}\left(1-e^{-\alpha_+(s) z}\right)\right.\nonumber\\
&+&\left.\frac{\gamma-\alpha_-(s)}{\alpha_-(s)}\left(1-e^{-\alpha_-(s) z}\right)\right],
\label{Exp_b}
\end{eqnarray}
and then
\begin{eqnarray}
\hat{\mathcal{P}}_{[0,b]}(s;x)
&=&\frac{1+\Lambda \hat{\mathcal{P}}_{[0,b]}(s;0)}{\Delta(s)}\nonumber\\
&\times&\left[\frac{\alpha_+(s)-\gamma}{\alpha_+(s)}\left(1-e^{-\alpha_+(s)(b-x)}\right)\right.\nonumber\\
&+&\left.\frac{\gamma-\alpha_-(s)}{\alpha_-(s)}\left(1-e^{-\alpha_-(s) (b-x)}\right)\right].
\label{Exp_c}
\end{eqnarray}
The value of $\hat{\mathcal{P}}_{[0,b]}(s;0)$ is obtained after demanding self-consistency to expression (\ref{Exp_c}) by letting $x=0$. 
We will not pursue further our analysis to detail the  explicit form of $\hat{\mathcal{P}}_{[0,b]}(s;0)$ since  Laplace inversion in the $s$--variable is not possible.   

\begin{figure}[hbtp]
{
\includegraphics[width=0.9\columnwidth,keepaspectratio=true]{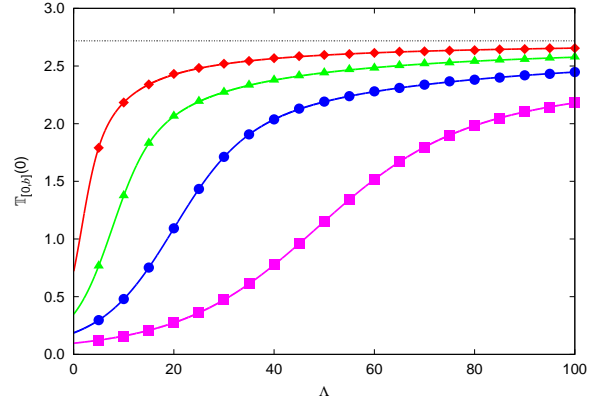}
}
\caption{(Color online) Sample representations of $\TT_{[0,b]}(0)$ in terms of $\Lambda$ for different choices of $\Gamma$. The solid lines were obtained by direct representation of the analytical result, whereas the points were computed through numerical simulation for: $\Gamma=1.0$,  (red) diamonds; $\Gamma=2.5$,  (green) triangles; $\Gamma=5.0$, (blue) circles; and $\Gamma=10.0$, (magenta) squares. The rest of parameters were $b=1.0$, $\lambda=1.0$, and $\gamma=1.0$.
}
\label{Fig_TT0}
\end{figure}

We will concentrate instead in the $s=0$ case, which leads to the MET, $\TT_{[0,b]}(x)$, as we have previously stated. From Eq.~(\ref{Exp_c}) we get
\begin{eqnarray}
\TT_{[0,b]}(x)
&=&\frac{1+\Lambda \TT_{[0,b]}(0)}{\Gamma\times(\alpha_+-\alpha_-)}\left[\frac{\alpha_+-\gamma}{\alpha_+}\left(1-e^{-\alpha_+(b-x)}\right)\right.\nonumber\\
&+&\left.\frac{\gamma-\alpha_-}{\alpha_-}\left(1-e^{-\alpha_-(b-x)}\right)\right].
\label{Exp_TTa}
\end{eqnarray}
The value of $\TT_{[0,b]}(0)$ must satisfy
\begin{eqnarray}
\TT_{[0,b]}(0)
&=&\frac{1+\Lambda \TT_{[0,b]}(0)}{\Gamma\times(\alpha_+-\alpha_-)}\left[\frac{\alpha_+-\gamma}{\alpha_+}\left(1-e^{-\alpha_+b}\right)\right.\nonumber\\
&+&\left.\frac{\gamma-\alpha_-}{\alpha_-}\left(1-e^{-\alpha_-b}\right)\right],
\label{Exp_TT0a}
\end{eqnarray}
that is, 
\begin{eqnarray}
\TT_{[0,b]}(0)=
\frac{1}{\Lambda}\left[\frac{\alpha_+-\alpha_-}{\left(\frac{\Lambda}{\Gamma} -\alpha_-\right)e^{-\alpha_+b}+\left(\alpha_+-\frac{\Lambda}{\Gamma}\right)e^{-\alpha_-b}} -1 \right], 
\nonumber \\
\label{Exp_TT0b}
\end{eqnarray}
which is always finite since
\begin{equation*}
\alpha_+>\frac{\Lambda}{\Gamma}>\alpha_-.
\end{equation*}
We can find in Fig.~\ref{Fig_TT0} some representations of $\TT_{[0,b]}(0)$ where we can observe the opposite effect of changing $\Lambda$ and $\Gamma$. 

\begin{figure}[hbtp]
{
\includegraphics[width=0.9\columnwidth,keepaspectratio=true]{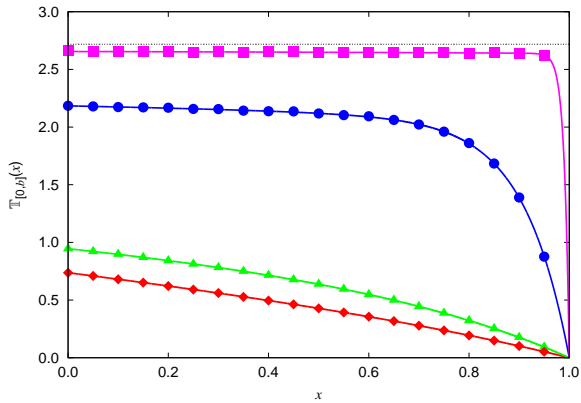}
}
\caption{(Color online) Sample representations of $\TT_{[0,b]}(x)$ for different values of $\Lambda$. The solid lines were obtained by direct representation of the analytical result, whereas the points were computed through numerical simulation for: $\Lambda=0.1$, (red) diamonds; $\Lambda=1.0$,  (green) triangles, $\Lambda=10.0$, (blue) circles; and finally $\Lambda=100.0$, (magenta) squares. The rest of the parameters were $b=1.0$, $\Gamma=1.0$, $\lambda=1.0$, and $\gamma=1.0$.
\label{Fig_TTx}
}
\end{figure}

Now, we combine Eqs.~(\ref{Exp_TTa}) and~(\ref{Exp_TT0b}) to obtain finally
\begin{eqnarray}
\TT_{[0,b]}(x)&=&
\frac{1}{(\Lambda-\Gamma\alpha_-)e^{-\alpha_+b}+(\Gamma\alpha_+-\Lambda)e^{-\alpha_-b}}\nonumber \\
&\times&
\left[\frac{\alpha_+-\gamma}{\alpha_+}\left(1-e^{-\alpha_+(b-x)}\right)\right.\nonumber\\
&+&\left.\frac{\gamma-\alpha_-}{\alpha_-}\left(1-e^{-\alpha_-(b-x)}\right)\right].
\label{Exp_TTb}
\end{eqnarray}
Figure~\ref{Fig_TTx}  shows how $\TT_{[0,b]}(x)$ depends on the starting point $x$, for different intensities of the reset mechanism.  We can observe how when the reset events are sparse $\TT_{[0,b]}(x)$ decreases gently, almost linearly, as a direct consequence of the constant drift term. Reciprocally, when there is large reset activity, the escape time is nearly insensitive to the value of $x$.  This means that in this regime the main possibility  for the process to leave the interval $[0,b]$  comes from the jump mechanism.

Figures~\ref{Fig_TT0} and~\ref{Fig_TTx} also include numerical estimations of the MET for a discrete set of values of the free parameter, $\Lambda$ and $x$ respectively. Once again we have used a Monte Carlo method in which $1,000,000$ different realizations of the process were averaged to produce the reported results. Note that in this case, since the estimation of the MET is a problem with a built-in stop mechanism, no time cut-off is necessary in the simulation of the process.

We conclude this section by considering some mathematical limits on Eq.~(\ref{Exp_TTb}). This will allow us to explore more deeply the limiting properties exposed in Figs.~\ref{Fig_TT0} and~\ref{Fig_TTx}. Letting $\Lambda \to 0$ no reset exists and we recover the reported result~\cite{VM10,VM11}:
\begin{eqnarray}
\lim_{\Lambda \to 0} \TT_{[0,b]}(x)
&=&\frac{\gamma }{\lambda +\Gamma \gamma}(b-x)\nonumber\\
&+&\frac{\lambda}{(\lambda +\Gamma \gamma)^2}\left[1-e^{-\frac{\lambda +\Gamma \gamma}{\Gamma}(b-x)} \right].
\label{Exp_TTL_0}
\end{eqnarray}
Letting $\Lambda \to \infty$ we find
\begin{eqnarray}
\lim_{\Lambda \to \infty} \TT_{[0,b]}(x)
&=&\frac{e^{\gamma b}}{\lambda}. 
\label{Exp_TTL_infty}
\end{eqnarray}
There exists an alternative, more physical derivation of this result by noting  that when $\Lambda \to \infty$  the system is constantly redirected to $0$, and exit can only happen when a jump of magnitude greater than $b$ occurs. Hence the MET cannot depend on initial position $x$ nor in  the drift parameter.
Furthermore one must also have
\begin{eqnarray*} 
\lim_{\Lambda \to \infty} \TT_{[0,b]}(x) &=&
\sum_{n=1}^{\infty} \EE\left[t_n-t_0|J_1,\dots,  J_{n-1}\le b, J_n>b\right]\\
&=&\sum_{n=1}^{\infty} n\EE\left[\tau_n\right]\PP\left\{ J_1,\dots, J_{n-1}\le b, J_n>b\right\} \\
&=&\sum_{n=1}^{\infty} \frac{n}{\lambda} (1- e^{-\gamma b})^{n-1}e^{-\gamma b}= \frac{e^{ \gamma b}}{\lambda},
\end{eqnarray*}
and the former result is recovered. This constant value is depicted in both Figs.~\ref{Fig_TT0} and~\ref{Fig_TTx} by a horizontal dotted (black) line.

We can also explore the limits involving the magnitude of the drift:
\begin{equation}
\lim_{\Gamma \to 0} \TT_{[0,b]}(x)
=e^{\frac{\Lambda \gamma}{\lambda +\Lambda}b}\left\{\frac{1}{\lambda}+\frac{1}{\Lambda }\left[1-e^{-\frac{\Lambda \gamma}{\lambda +\Lambda}(b-x)} \right]\right\},
\label{Exp_TTG0}
\end{equation}
and
\begin{equation}
\lim_{\Gamma \to \infty} \TT_{[0,b]}(x)=0.
\label{Exp_TTG8}
\end{equation}

\section{Conclusions}
\label{Sec_conclusions}

In this paper we have analyzed in detail the main statistical magnitudes of a stochastic process that experiences random reset events whereupon the system returns to the initial point. As the basic stochastic process we take a monotonous continuous-time random walk with a constant drift: the process increases between the reset events, either by the effect of the random jumps, or by the action of the deterministic drift. We find that the system that results from the balance of these 
opposite effects has interesting properties, like the existence of a stationary transition probability density function for any choice of jump density and drift strength, or the faculty of the model to reproduce power-law-like behavior for a certain range in the parameter space. If the jump density is exponential the transition density can be computed explicitly. The long time limit yields a heavy-tailed density with a simple power-law exponent. Similarly, exit times of the model can also be determined.  Analytic results are  found to be in a remarkable agreement with the numerical observations obtained using Monte Carlo analysis to average over a large number of realizations.

Finally note that we have concentrated our efforts in the case in which both the waiting times between successive reset events 
and consecutive jumps have a Poissonian origin, in the same line of the previous work by Manrubia and Zanette~\cite{MZ99}. The inclusion of alternative innovation mechanisms is deferred to a future publication.


\begin{acknowledgments}
The authors acknowledge support from the former Spanish Ministerio de Ciencia e Innovaci\'on under Contracts No. FIS2009-09689 (MM) and MTM2009-09676 (JV), and from Generalitat de Catalunya, Contract No. 2009SGR417 (MM).
\end{acknowledgments}

\appendix

\section{The general propagator}
\label{App_ansatz}
Here we analyze in more detail how $p(x,\tau;x_0)$ depend on $x_0$.  This will allow us to solve Eq.~\eqref{LTPDF} and, eventually, to recover Eq.~\eqref{PDF_final}.

Assume  first that $x<x_0+\Gamma \tau$. In that case the process must at least experience one reset before $\tau$ since otherwise it could not have reached $x$ starting from $x_0$. Since the reset event removes any memory on past locations of the process, we must have
\begin{equation*}
p(x,\tau;x_0)= g(x,\tau),
\end{equation*}
independent of $x_0$. When $x\geq x_0+\Gamma \tau$ there is {\it in addition\/} a possibility that the process goes from $x_0$ to $x$ directly, with no reset event in the meanwhile,
\begin{equation*}
p(x,\tau;x_0)= g(x,\tau)+Q(x,\tau;x_0).
\end{equation*}
Therefore, $Q(x,\tau;x_0)$ corresponds to the possibility that the process reaches $x$ starting from $x_0$ by the combined action of the random jumps and the constant drift. The latter contribution will affect the stochastic process in a deterministic way, by decreasing the {\it effective distance\/} that the process must travel by means of the random jumps alone. Since the jumping mechanism is space-homogeneous we must have that, for a given time interval $\tau$, $Q(x,\tau;x_0)$ {\it must depend on $x$ and $x_0$ only via the precise combination of variables\/} $x-x_0-\Gamma \tau$:
\begin{equation*}
Q(x,\tau;x_0)=q(x-x_0-\Gamma \tau,\tau).
\end{equation*}
Thus, summing up, $p(x,\tau;x_0)$ must have the structure
\begin{equation}
p(x,\tau;x_0)\equiv g(x,\tau)+q(x-x_0-\Gamma \tau,\tau)\theta(x-x_0-\Gamma \tau),
\label{x0_dependence}
\end{equation}
for some functions $g(z,\tau)$ and $q(z,\tau)$. Note that the full process with drift cannot be recovered from the driftless process by the Galilean transformation $x\rightarrow x-\Gamma \tau$.

We can still gain more insight on $g(x,\tau)$. This function although related with $p(x,\tau;0)$ is not the same object. In $g(x,\tau)$ at least one reset event has occurred within the time interval $\tau$. This means that we know that the time of the last reset event is positive, $0\leq T^*(\tau)\leq \tau$, and from that point the process has to reach $x$. Therefore, if we define $\tau^*\equiv \tau-T^*(\tau)$, $0\leq\tau^*\leq \tau$, it follows with a similar reasoning to that which led to Eq.~(\ref{x0_dependence}) that we must demand $g(x,\tau)$ to conform to the following structure:
\begin{equation*}
g(x,\tau)=\int_0^{\tau}d\tau^* K(x-\Gamma \tau^*,\tau^*)\theta(x-\Gamma \tau^*).
\end{equation*}
Here the kernel $K(x-\Gamma \tau^*,\tau^*)$ includes the probability of reaching $x$ from the origin after time $\tau^*$ {\it as well as\/} the fact that the last reset event was at $T^*(\tau)=\tau-\tau^*$; the step function simply ensures that $x$ is still accessible. Based on the Poissonian nature of the reset events we could argue that indeed $K(x-\Gamma \tau^*,\tau^*)=\Lambda q(x-\Gamma \tau^*,\tau^*)$, but that is not a necessary ingredient at this point: we will recover this result as a byproduct of the problem-solving procedure.

This procedure begins with the computation of the double Laplace transform of our {\it ansatz\/} for $p(x,\tau;x_0)$,
\begin{eqnarray}
p(x,\tau;x_0)&=&\int_0^\tau d\tau^* K(x-\Gamma \tau^*,\tau^*)\theta(x-\Gamma \tau^*)\nonumber \\
&+&q(x-x_0-\Gamma \tau,\tau)\theta(x-x_0-\Gamma \tau), 
\label{Ansatz_p}
\end{eqnarray}
which reads
\begin{eqnarray}
\hat{\tilde{p}}(\omega,s;x_0)&=&\frac{1}{s} \hat{\tilde{K}}(\omega,s+\Gamma \omega)\nonumber \\ &+&
\hat{\tilde{q}}(\omega,s+\Gamma \omega)e^{-\omega x_0}. 
\label{Ansatz_Lp}
\end{eqnarray}
Inserting this result into the right-hand side of Eq.~(\ref{LTPDF}) we obtain
\begin{eqnarray}
\hat{\tilde{p}}(\omega,s;x_0)
&=&\frac{1}{\lambda +\Lambda+s+\Gamma \omega}e^{-\omega x_0} \nonumber\\
&+&\frac{\Lambda}{\lambda +\Lambda+s}\left[\frac{1}{s} \hat{\tilde{K}}(\omega,s+\Gamma \omega)+
\hat{\tilde{q}}(\omega,s+\Gamma \omega)\right]
\nonumber\\
&+&\frac{\lambda}{\lambda +\Lambda+s+\Gamma \omega} \tilde{h}(\omega)\hat{\tilde{q}}(\omega,s+\Gamma \omega) e^{-\omega x_0}\nonumber\\ 
&+&\frac{\lambda}{\lambda +\Lambda+s}\frac{1}{s}\hat{\tilde{K}}(\omega,s+\Gamma \omega),
\label{Ansatz_LTPDF}
\end{eqnarray}
where 
\begin{equation*}
\tilde{h}(\omega)\equiv\int_0^{\infty}du h(u) e^{-\omega u}.
\end{equation*}
Let us compare the output in Eq.~(\ref{Ansatz_LTPDF})  with Eq.~(\ref{Ansatz_Lp}): On the one hand equating the terms that contains the factor $e^{-\omega x_0}$ we obtain
\begin{eqnarray}
\hat{\tilde{q}}(\omega,s+\Gamma \omega)
&=&\frac{1}{\lambda +\Lambda+s+\Gamma \omega} \nonumber\\
&+&\frac{\lambda}{\lambda +\Lambda+s+\Gamma \omega}
\tilde{h}(\omega) \hat{\tilde{q}}(\omega,s+\Gamma \omega),
\nonumber\\
\end{eqnarray}
thus we find
\begin{equation}
\hat{\tilde{q}}(\omega,s)=\frac{1}{\lambda\left[1-\tilde{h}(\omega)\right] +\Lambda+s}.
\label{q_sol}
\end{equation}
On the other hand considering the terms which are independent from $x_0$ we get
\begin{eqnarray}
\frac{1}{s} \hat{\tilde{K}}(\omega,s+\Gamma \omega)
&=&\frac{1}{\lambda +\Lambda+s}\bigg[ \frac{\Lambda+\lambda}{s}\hat{\tilde{K}}(\omega,s+\Gamma \omega) 
\nonumber\\
&+&\Lambda \hat{\tilde{q}}(\omega,s+\Gamma \omega) \bigg],
\end{eqnarray}
and hence
\begin{equation}
\hat{\tilde{K}}(\omega,s)=\Lambda \hat{\tilde{q}}(\omega,s),
\label{K_sol}
\end{equation}
as we have already anticipated. Therefore, in the Laplace domain the solution of the problem reads:
\begin{equation*}
\hat{\tilde{p}}(\omega,s;x_0)=\left[\frac{\Lambda}{s}+e^{-\omega x_0}\right] \frac{1}{\lambda\left[1-\tilde{h}(\omega)\right] +\Lambda+s+\Gamma \omega}.
\end{equation*}

Let us analyze the structure of the above equation. Note how it can be expressed as:
\begin{equation}
\hat{\tilde{p}}(\omega,s;x_0)=\frac{\Lambda}{s}\hat{\tilde{p}}_0(\omega,s+\Lambda;0)+\hat{\tilde{p}}_0(\omega,s+\Lambda;x_0),
\label{PDF_semi_final}
\end{equation}
with
\begin{equation*}
\hat{\tilde{p}}_0(\omega,s;x_0)\equiv\lim_{\Lambda\to 0}\hat{\tilde{p}}(\omega,s;x_0)=\frac{e^{-\omega x_0}}{\lambda\left[1-\tilde{h}(\omega)\right] +\Gamma \omega + s}.
\end{equation*}
The interpretation of $p_0(x,\tau;x_0)$ is direct: it is the transition probability of the process between $x_0$ and $x$ when the reset mechanism is disconnected. 
Therefore, by appealing to well-known properties of Laplace transforms|see Eqs.~\eqref{I_s_shift} and~\eqref{I_int_tau} below|, expression~\eqref{PDF_semi_final} leads to Eq.~\eqref{PDF_final}:
\begin{eqnarray*}
p(x,\tau;x_0)&=&\int_0^\tau d\tau^* \Lambda e^{-\Lambda \tau^*} p_0(x,\tau^*;0)\nonumber \\
&+&p_0(x,\tau;x_0)e^{-\Lambda \tau}.
\end{eqnarray*}
Note how we have used $\tau^*$ as the integration variable. This fact is not incidental at all: $\tau^*$ was defined above as the time period comprised between the last reset and $\tau$, and due to the Poissonian nature of the reset events, this interval is exponentially distributed with intensity $\Lambda$. The interpretation of  Eq.~\eqref{PDF_final} is the following: {\it the probability of going from $x_0$ to $x$ is the sum of that probability  when no reset has taken place, plus the probability of reaching $x$ after the   last reset event\/}. This probabilistic construction is valid because the propagator does not depend on the whole path of the process.

\section{Markovianity, invariance and irreducibility}
\label{App_Markov}

This appendix contains the proofs of several probabilistic properties which have been used in the text. We begin with the Markov property. It is well known~\cite{Lefever,kt81} that a time-homogeneous stochastic process $X(\tau)$ is a Markov process if and only if its transition probability satisfies the Chapman-Kolmogorov equation (CKE):  
\begin{equation}
p(x,\tau+\bar{\tau};x_0)=\int_0^{\infty}d\bar{x} p(x,\tau;\bar{x}) p(\bar{x},\bar{\tau};x_0),
\label{CKE}
\end{equation}
for any $\bar{\tau}$, $\tau$.  We also note that if  $X(\tau)$ is Markov then so it is $Y(\tau)$. 

A simple  way to prove \eqref{CKE} is to use the fact that the reset-free process is Markovian and that propagators of the process with and without the reset mechanism, namely
$p(x,\tau;x_0)$  and $p_0(x,\tau;x_0)$, are related by Eq.~\eqref{PDF_final}. Indeed it implies at once that the right hand side of Eq.~\eqref{CKE} can be written as a sum of several factors:
\begin{equation*}
p(x,\tau+\bar{\tau};x_0)=F_1+F_2+F_3+F_4.
\end{equation*}
Here
\begin{eqnarray}
F_2&\equiv&\int_0^{\infty}d\bar{x} p_0(x,\tau;\bar{x})  e^{-\Lambda \tau} \int_0^{\bar{\tau}} d\tau^* \Lambda e^{-\Lambda \tau^*} p_0(\bar{x},\tau^*;0)\nonumber\\
&=&\int_0^{\bar{\tau}} d\tau^*\Lambda  e^{-\Lambda (\tau+\tau^*)}p_0(x,\tau+\tau^*;0)\nonumber\\
&=& \int_{\tau}^{\tau+\bar{\tau}} d\tau^*\Lambda e^{-\Lambda \tau^*}p_0(x,\tau^*;0),
\label{P2}
\end{eqnarray}
\begin{eqnarray}
F_3&\equiv&\int_0^{\infty}d\bar{x} \int_0^\tau d\tau^* \Lambda e^{-\Lambda \tau^*} p_0(x,\tau^*;0) p_0(\bar{x},\bar{\tau};x_0)e^{-\Lambda \bar{\tau}}\nonumber\\
&=&\int_0^\tau d\tau^* \Lambda e^{-\Lambda (\bar{\tau}+\tau^*)} p_0(x,\tau^*;0),
\label{P3}
\end{eqnarray}
and
\begin{eqnarray}
F_4&\equiv&\int_0^{\infty}d\bar{x} \int_0^\tau d\tau^* \Lambda e^{-\Lambda \tau^*} p_0(x,\tau^*;0)\nonumber\\
&\times&\int_0^{\bar{\tau}} d\bar{\tau}^* \Lambda e^{-\Lambda \bar{\tau}^*} p_0(\bar{x},\bar{\tau}^*;0)\nonumber\\
&=&\int_0^\tau d\tau^* \Lambda e^{-\Lambda \tau^*} p_0(x,\tau^*;0) \int_0^{\bar{\tau}}d\bar{\tau}^* \Lambda e^{-\Lambda \bar{\tau}^*} \nonumber\\
&=&\int_0^\tau d\tau^* \Lambda e^{-\Lambda \tau^*} p_0(x,\tau^*;0) \left(1- e^{-\Lambda \bar{\tau}}\right). 
\label{P4}
\end{eqnarray}
Hence these three factors add to
\begin{eqnarray}
F_2+F_3+F_4&=&\int_{\tau}^{\tau+\bar{\tau}} d\tau^*\Lambda e^{-\Lambda \tau^*}p_0(x,\tau^*;0)\nonumber\\
&+&\int_0^\tau d\tau^* \Lambda e^{-\Lambda \tau^*} p_0(x,\tau^*;0)\nonumber\\
&=&\int_{0}^{\tau+\bar{\tau}} d\tau^*\Lambda e^{-\Lambda \tau^*}p_0(x,\tau^*;0).
\label{P234}
\end{eqnarray}
We now consider the remaining term
\begin{eqnarray}
F_1&\equiv&\int_0^{\infty}d\bar{x} p_0(\bar{x},\bar{\tau};x_0)p_0(\bar{x},\bar{\tau};x_0)e^{-\Lambda (\tau+\bar{\tau})}\nonumber\\
&=&p_0(x,\tau+\bar{\tau};x_0) e^{-\Lambda (\tau+\bar{\tau})},
\label{P1}
\end{eqnarray}
where we have used that the process without resets is Markovian and hence $p_0(x,\tau;x_0)$ {\it must satisfy}  its own CKE similar to~\eqref{CKE}:
\begin{equation}
p_0(x,\tau+\bar{\tau};x_0)=\int_0^{\infty}d\bar{x} p_0(x,\tau;\bar{x}) p_0(\bar{x},\bar{\tau};x_0).
\label{CKE0}
\end{equation}
Finally, using again Eq.~\eqref{PDF_final} we have
\begin{eqnarray}
F_1+F_2+F_3+F_4&=&p_0(x,\tau+\bar{\tau};x_0) e^{-\Lambda (\tau+\bar{\tau})}\nonumber\\
&+&\int_{0}^{\tau+\bar{\tau}} d\tau^*\Lambda e^{-\Lambda \tau^*}p_0(x,\tau^*;0)\nonumber\\
&=&p(x,\tau+\bar{\tau};x_0).
\label{P1234}
\end{eqnarray}
Equation~\eqref{CKE} now follows.

Next, by letting $\bar{\tau}\to\infty$, Eq.~\eqref{CKE} implies that
\begin{equation*}
p(x)=\int_0^{\infty}d\bar{x}  p(x,\tau;\bar{x}) p(\bar{x}),
\end{equation*}
where we remind that $p(x)\equiv \lim_ {\tau\to\infty}  p(x,\tau;x_0)$, is independent of $x_0$, the initial state. This means that if we start with an initial distribution $p(\cdot)$ then {\it  for all time the density of the process is also\/} $p(\cdot)$. Thus the limit distribution is {\it invariant under the dynamics\/}.

Finally, we prove that the system is irreducible, namely, that there is positive probability to visit any region $A\equiv [x,x+\epsilon]$, $x\geq0$, starting from $x_0\geq 0$.  An analysis of the sample path trajectories along with the fact that $J_n$ has a continuous density function $h(\cdot)$ will convince the reader that this is indeed the case. For example, if $x_0< x$ then $X(\tau)$ can reach $A$ with just one jump. Consequently the probability $P(A,\tau;x_0)$ to visit $A$ before time $\tau$ satisfies the lower bound
\begin{equation*}
P(A,\tau;x_0) \geq e^{-\Lambda \tau}\left(1-e^{-\lambda \tau}\right)\int_{x-x_0}^{x+\epsilon-x_0}h(\bar{x})d\bar{x}>0.
\end{equation*}
The case $x+\epsilon<x_0$ can be handled similarly with the only proviso that a reset must occur before $\tau$  whereupon $A$ may be visited via jumps:
\begin{eqnarray*}
 P(A,\tau) &\geq&  \left[\frac{\Lambda}{\Lambda-\lambda}\left(1-e^{-\lambda \tau}\right)+\frac{\lambda}{\lambda-\Lambda}\left(1-e^{-\Lambda \tau}\right)\right] \\
&\times& \int_{x}^{x+\epsilon}h(\bar{x})d\bar{x}>0.
\end{eqnarray*}

\section{Derivation of Eq.~\eqref{p_x_tau}}
\label{app_p_x_tau}

This appendix shows the how the basic result Eq.~\eqref{p_x_tau} can be recovered by Laplace inversion of Eq.~\eqref{p_h_exp_s}.

We first introduce some notational conventions and list two basic properties of the Laplace transform that are used below. For the sake of brevity, we denote by $\mathcal{L}$ and $\mathcal{L}^{-1}$ the direct and inverse Laplace transform operator respectively:
\begin{eqnarray*}
\hat{f}(s)&\equiv& \mathcal{L}[f(\tau),\tau,s],\\
f(\tau)&\equiv&\mathcal{L}^{-1}[\hat{f}(s),s,\tau].
\end{eqnarray*}
The first property is the so-call {\it frequency shifting\/}
\begin{equation}
\mathcal{L}\left[f(\tau)e^{-s_0\tau},\tau,s\right]=\hat{f}(s+s_0),
\label{s_shift}
\end{equation}
which can be alternatively expressed as
\begin{equation}
\mathcal{L}^{-1}\left[\hat{f}(s+s_0),s,\tau\right]=\mathcal{L}^{-1}\left[\hat{f}(s),s,\tau\right]e^{-s_0\tau}.
\label{I_s_shift}
\end{equation}
The second property concerns the interplay between integration and Laplace transform:
\begin{equation}
\mathcal{L}\left[\int_0^{\tau} f(\bar{\tau})d\bar{\tau},\tau,s\right]=\frac{1}{s}\hat{f}(s),
\label{int_tau}
\end{equation}
or
\begin{equation}
\mathcal{L}^{-1}\left[\frac{1}{s}\hat{f}(s),s,\tau\right]=\int_0^{\tau} d\bar{\tau}\mathcal{L}^{-1}\left[\hat{f}(s),s,\bar{\tau}\right].
\label{I_int_tau}
\end{equation}

Thus, we want to compute
\begin{equation*}
p(x,\tau;x_0)\equiv\mathcal{L}^{-1}[\hat{p}(x,s;x_0),s,\tau],
\end{equation*}
for
\begin{eqnarray*}
\hat{p}(x,s;x_0)&=&\frac{1}{\lambda+\Lambda+s}\left[\frac{\Lambda}{s}\delta(x)+\delta(x-x_0)\right]\\
&+&\frac{\Lambda }{s}\frac{\gamma \lambda}{\left(\lambda+\Lambda+s\right)^2}e^{-\frac{(\Lambda+s)\gamma x}{\lambda+\Lambda+s}}
\\
&+&\frac{\gamma\lambda}{\left(\lambda+\Lambda+s\right)^2} e^{-\frac{\Lambda+s}{\lambda+\Lambda+s}\gamma (x-x_0)}\theta(x-x_0).
\end{eqnarray*}
Since the Laplace operator is linear, we can write with obvious notation
\begin{eqnarray}
p(x,\tau;x_0)=L_1+L_2+L_3+L_4.
\label{L_def}
\end{eqnarray}

Let us begin with the simplest term:
\begin{equation}
L_2\equiv\mathcal{L}^{-1}\left[\frac{1}{\lambda+\Lambda+s},s,\tau\right] \delta(x-x_0)=e^{-(\lambda+\Lambda)\tau}\delta(x-x_0),
\end{equation}
where we have used the property~\eqref{I_s_shift} for $s_0=\lambda+\Lambda$ and the fact that 
\begin{equation}
\mathcal{L}^{-1}\left[\frac{1}{s},s,\tau\right]=1.
\label{one_over_s}
\end{equation}

The next term $L_1$ is very similar to $L_2$ using partial fraction decomposition:
\begin{eqnarray}
L_1&\equiv&
\mathcal{L}^{-1}\left[\frac{\Lambda}{s}\frac{1}{\lambda+\Lambda+s},s,\tau\right] \delta(x)\nonumber\\
&=&\frac{\Lambda}{\lambda+\Lambda}\mathcal{L}^{-1}\left[\frac{1}{s}-\frac{1}{\lambda+\Lambda+s},s,\tau\right]\delta(x)\nonumber\\
&=&\frac{\Lambda}{\lambda + \Lambda}\left[1-e^{-(\lambda+\Lambda)\tau}\right]\delta(x).
\end{eqnarray}

The fourth term reads 
\begin{eqnarray}
L_4&\equiv&
\mathcal{L}^{-1}\left[\frac{\gamma\lambda}{\left(\lambda+\Lambda+s\right)^2} e^{-\frac{\Lambda+s}{\lambda+\Lambda+s}\gamma (x-x_0)},s,\tau\right] \theta(x-x_0)\nonumber\\
&=&\gamma\lambda e^{-(\lambda+\Lambda)\tau}\mathcal{L}^{-1}\left[\frac{1}{s^2} e^{-\frac{s-\lambda}{s}\gamma (x-x_0)},s,\tau\right] \theta(x-x_0)\nonumber\\
&=&\gamma\lambda e^{-(\lambda+\Lambda)\tau-\gamma (x-x_0)}\nonumber\\
&\times&\mathcal{L}^{-1}\left[\frac{1}{s^2} e^{\frac{\lambda}{s}\gamma (x-x_0)},s,\tau\right]\theta(x-x_0) \nonumber \\
&=&e^{-(\lambda+\Lambda)\tau-\gamma (x-x_0)}\sqrt{\frac{\gamma \lambda \tau}{x-x_0}} \nonumber\\
&\times& I_1\left(2\sqrt{\lambda \tau\gamma(x-x_0)}\right)\theta(x-x_0).
\label{L_4} 
\end{eqnarray}
where we have used again Eq.~\eqref{I_s_shift} and also that:
\begin{eqnarray}
\mathcal{L}^{-1}\left[\frac{1}{s^2} e^{\frac{c}{s}},s,\tau\right]=\sqrt{\frac{\tau}{c}}  I_1\left(2\sqrt{c \tau}\right),
\label{L_Bessel}
\end{eqnarray}
see~\cite{RK66}. Here $I_1(\cdot)$ is the modified Bessel function of order 1. 

Only the term $L_3$ remains. It can be handled by using Eqs.~\eqref{I_int_tau},~\eqref{I_s_shift} and~\eqref{L_Bessel}
consecutively:
\begin{eqnarray}
L_3&\equiv&
\mathcal{L}^{-1}\left[\frac{\Lambda }{s}\frac{\gamma \lambda}{\left(\lambda+\Lambda+s\right)^2}e^{-\frac{(\Lambda+s)\gamma x}{\lambda+\Lambda+s}},s,\tau\right] \nonumber\\
&=&\Lambda\gamma \lambda \int_0^\tau d\bar{\tau} \mathcal{L}^{-1}\left[\frac{1}{\left(\lambda+\Lambda+s\right)^2}e^{-\frac{(\Lambda+s)\gamma x}{\lambda+\Lambda+s}},s,\bar{\tau}\right]  \nonumber\\
&=&\Lambda\gamma \lambda \int_0^\tau d\bar{\tau} \mathcal{L}^{-1}\left[\frac{1}{s^2}e^{\frac{\lambda\gamma x}{s}},s,\bar{\tau}\right] e^{-(\lambda+\Lambda)\bar{\tau}-\gamma x}\nonumber\\
&=&\Lambda\int_0^\tau d\bar{\tau} e^{-(\lambda+\Lambda)\bar{\tau}-\gamma x}\sqrt{\frac{\gamma \lambda \bar{\tau}}{x}} I_1\left(2\sqrt{\lambda \bar{\tau}\gamma x}\right).\nonumber\\
\label{L_3} 
\end{eqnarray} 
 
Putting together all the terms, Eq.~\eqref{p_x_tau} follows.

\section{From survival probabilities to mean exit times}
\label{App_s_is_0}
According to the results of Sec.~\ref{Sec_extreme} and Eq.~\eqref{survival}, the mean exit time $\TT_{[a,b]}(x_0)$ may be computed as follows
\begin{eqnarray*}
\TT_{[a,b]}(x_0)&=&\int_0^\infty \tau d \PP\left\{\mathcal{T}\leq\tau\right\}=-\int_0^\infty \tau d\mathcal{P}_{[a,b]}(\tau;x_0)\\
&=&\int_0^\infty \mathcal{P}_{[a,b]}(\tau;x_0)d\tau,
\end{eqnarray*}
where in the last equality we have assumed that $\mathcal{P}_{[a,b]}(\tau;x_0)$ decays appropriately as $\tau\to\infty$ to perform safely integration by parts. Now, note that by definition
\begin{eqnarray*}
\hat{\mathcal{P}}_{[a,b]}(s=0;x_0)&=&\left.\int_0^\infty \mathcal{P}_{[a,b]}(\tau;x_0)e^{-s\tau
}d\tau\right|_{s=0}\\
&=&\int_0^\infty \mathcal{P}_{[a,b]}(\tau;x_0)d\tau,
\end{eqnarray*}
and thus finally
\begin{equation}
\TT_{[a,b]}(x_0)=\hat{\mathcal{P}}_{[a,b]}(s=0;x_0).
\label{TT_LSP}
\end{equation}


\begin{thebibliography}{00}

\bibitem{Lefever} \Book{W. Horsthemke and R. Lefever}{2007}{Noise-Induced Transitions: Theory and Applications in Physics, Chemistry and Biology}{Springer}{Amsterdam} %

\bibitem{kt81} \Book{S. Karlin and H. Taylor}{1981}{A second course in stochastic processes}{Academic Press}{New York} %

\bibitem{Vtmp} \Journal{J. Villarroel}{2005}{Killed random processes and heat kernels}{Theor. Math. Phys.}{144}{1238}{1245} 

\bibitem{MZ99} \Journal{S. C. Manrubia and D. H. Zanette}{1999}{Stochastic multiplicative processes with reset events}{Phys. Rev. E}{59}{4945}{4948} 

\bibitem{EM11a} \Journal{M. R. Evans and S. N. Majumdar}{2011}{Diffusion with Stochastic Resetting}{Phys. Rev. Lett.}{106}{160601}{} 

\bibitem{EM11b} \Journal{M. R. Evans and S. N. Majumdar}{2011}{Diffusion with optimal resetting}{J. Phys. A: Math. Theor.}{44}{435001}{} 



\bibitem{Man} \Journal{B. Mandelbrot}{1963}{The Variation of Certain Speculative Prices}{J. Bus.}{36}{394}{419} 

\bibitem{Ma} \Journal{R. N. Mantegna and H. E. Stanley}{1995}{Scaling behavior in the dynamics of an economic index}{Nature}{376}{46}{49} 

\bibitem{Stan}\Journal{H. A. Makse, S. Havlin, and H. E. Stanley}{1995}{Modelling urban growth patterns}{Nature}{377}{608}{612} 

\bibitem{Sorn}\Journal{D. Sornette}{1998}{Multiplicative processes and power laws}{Phys. Rev. E}{57}{4811}{4813} 

\bibitem{BS} \Book{N. N. Taleb}{2007}{The Black Swan: The Impact of the Highly Improbable}{Random House}{New York} %

\bibitem{Sch18} \Journal{W. Schottky}{1918}{Über spontane Stromschwankungen in verschiedenen Elektrizitätsleitern}{Ann. Phys.-Berlin}{362}{541}{567} %

\bibitem{Rice} \Journal{S. O. Rice}{1945}{Mathematical Analysis of Random Noise} {Bell Syst. Tech. J.}{24}{46}{156} 

\bibitem{MW65} \Journal{E. W. Montroll and G. H. Weiss}{1965}{Random Walks on Lattices, II}{J. Math. Phys.}{6}{167}{181} 

\bibitem{W94} \Book{G. H. Weiss}{1994}{Aspects and Applications of the Random Walk}{North-Holland}{Amsterdam} %

\bibitem{S74} \Journal{M. F. Shlesinger}{1974}{Asymptotic solutions of continuous-time random walks} {J. Stat. Phys.}{10}{421}{434} %

\bibitem{ms84} \Book{E. W. Montroll and M. F. Shlesinger}{1984}{Nonequilibrium
Phenomena II: From stochastics to hydrodynamics. {\rm In: J. L. Lebowitz, E. W. Montroll (Eds.), pp. 1-121}}{North-Holland}{Amsterdam} %

\bibitem{Hu} \Journal{B. D. Hughes,  E. W.  Montroll, and  M. F. Shlesinger}{1982}{Fractal random walks}{J. Stat. Phys.}{28}{111}{126} %

\bibitem{Weiss-porra} \Journal{G. H. Weiss, J. M. Porr\`a, and J. Masoliver}{1998}{Statistics of the depth probed by cw measurements of photons in a turbid medium}{Phys. Rev. E}{58}{6431}{6439} 

\bibitem{Margolin1} \Journal{G. Margolin and B. Berkowitz}{2002}{Spatial behavior of anomalous transport}{Phys. Rev. E}{65}{031101}{} %

\bibitem{hs02} \Journal{A. Helmstetter and D. Sornette}{2002}{Diffusion of epicenters of earthquake aftershocks, Omori's law, and generalized continuous-time random walk models}{Phys. Rev. E}{66}{061104}{} 

\bibitem{Me03} \Journal{M. S. Mega, P. Allegrini, P. Grigolini, V. Latora, and  L. Palatella}{2003}{Power-Law Time Distribution of Large Earthquakes}{Phys. Rev. Lett.}{90}{188501}{} 

\bibitem{Be} \Journal{B. Berkowitz and H. Scher}{1997}{Anomalous Transport in Random Fracture Networks}{Phys. Rev. Lett.}{79}{4038}{4041} 

\bibitem{Bo} \Journal{M. Bogu\~n\'a and  \'A. Corral}{1997}{Long-Tailed Trapping Times and Lévy Flights in a Self-Organized Critical Granular System}{Phys. Rev. Lett.}{78}{4950}{4953}


\bibitem{Gu} \Journal{E. Gudowska-Nowak and K. Weron}{2002}{Random walk models of electron tunneling in a fluctuating medium}{Phys. Rev. E}{65}{011103}{} 

\bibitem{Os71} \Journal{V. S. Oskanian and V. Yu. Terebizh}{1971}{}{Astrophysics}{7}{48}{54} %





\bibitem{mmw03} \Journal{J. Masoliver, M. Montero, and G. H. Weiss}{2003}{Continuous-time random-walk model for financial distributions}{Phys. Rev. E}{67}{021112}{} 

\bibitem{mmp05} \Journal{J. Masoliver, M. Montero, and J. Perell\'o}{2005}{Extreme times in financial markets}{Phys. Rev. E}{71}{056130}{} 

\bibitem{mmpw06} \Journal{J. Masoliver, M. Montero, J.  Perell\'o, and G. H. Weiss}{2006}{The continuous time random walk formalism in financial markets}{J. Econ. Behav. Organ.}{61}{577}{598} 

\bibitem{mpmlmm05} \Journal{M. Montero, J.  Perell\'o, J. Masoliver,  F. Lillo, S. Miccich\'{e}, and R. N. Mantegna}{2005}{Scaling and data collapse for the mean exit time of asset prices}{Phys. Rev. E}{72}{056101}{} 


\bibitem{s06} \Journal{E. Scalas}{2006}{The application of continuous-time random walks in finance and economics}{Physica A}{362}{225}{239}

\bibitem{Metzler} \Journal{R. Metzler and J. Klafter}{2000}{The random walk:s guide to anomalous diffusion: A fractional dynamics approach}{Phys. Rep.}{339}{1}{77} 

\bibitem{s04} \Journal{E. Scalas, R. Gorenflo, and  F. Mainardi}{2004}{Uncoupled continuous-time random walks: Solution and limiting behavior of the master equation}{Phys. Rev. E}{69}{011107}{} 

\bibitem{m07} \Journal{F. Mainardi, R. Gorenflo, and A. Vivoli}{2007}{Beyond the Poisson renewal process: A tutorial survey}{J. Comput. Appl. Math.}{205}{725}{735} 

\bibitem{Gerbershiu} \Journal{H. U. Gerber and E. S. W. Shiu}{1996}{Actuarial Bridges to Dynamic Hedging and Option Pricing}{Insur. Math. Econ.}{18} {183}{218} 

\bibitem{Dickson} \Journal{D. C. M . Dickson and C. Hipp}{2001}{On the time of ruin for Erlang(2) risk processes}{Insur. Math. Econ.}{29}{333}{344} 

\bibitem{lg04} \Journal{S. Li and J. Garrido}{2004}{On ruin for the Erlang(n) risk process}{Insur. Math. Econ.}{34}{391}{408} 

\bibitem{zyl10} \Journal{Z. Zhanga, H. Yanga, and S. Li}{2010}{The perturbed compound Poisson risk model with two-sided jumps}{J. Comput. Appl. Math.}{233}{1773}{1784} 

\bibitem{VM10}  \Journal{J. Villarroel and M. Montero}{2010}{On the effect of random inhomogeneities in Kerr media modelled by a nonlinear Schr\"odinger equation}{J. Phys. B: At. Mol. Opt. Phys.}{43}{135404}{} 

\bibitem{VM11}  \Journal{J. Villarroel and M. Montero}{2011} {On the Integrability of the Poisson Driven Stochastic Nonlinear Schr\"odinger Equations} {Stud. Appl. Math.}{127}{372}{393} 

\bibitem{MV10} \Journal{M. Montero and J. Villarroel}{2010}{Exit times in non-Markovian drifting continuous-time random-walk processes}{Phys. Rev. E}{82}{021102}{} 

\bibitem{ABCJS} \Journal{M. J. Ablowitz, G. Biondini, S. Chakravarty, R. B. Jenkins, and J. R. Sauer}{1996}{Four-wave mixing in wavelength-division-multiplexed soliton systems: damping and amplification}{Opt. Lett.}{21}{1646}{1648} 

\bibitem{RK66} \Book{G. E. Roberts and H. Kaufman}{1966}{Table of Laplace Transforms}{Saunders}{Philadelphia} %


\end{thebibliography}
\end{document}